\documentclass[twocolumn]{aastex63}
\usepackage{url}
\accepted{July 1, 2020}
\submitjournal{ApJ}

\shorttitle{Luminosity Function of protocluster galaxies}
\shortauthors{Ito et al.}

\begin{document}

\title{The UV Luminosity Function of Protocluster Galaxies at $z\sim4$: the Bright-end Excess and the Enhanced Star Formation Rate Density}

\correspondingauthor{Kei Ito}
\email{kei.ito@grad.nao.ac.jp}

\author[0000-0002-9453-0381]{Kei Ito}
\affiliation{Department of Astronomical Science, The Graduate University for Advanced Studies, SOKENDAI, Mitaka, Tokyo, 181-8588, Japan} 
\affiliation{National Astronomical Observatory of Japan, Mitaka, Tokyo, 181-8588, Japan}
\affiliation{Department of Astronomy, School of Science, The University of Tokyo, 7-3-1 Hongo, Bunkyo-ku, Tokyo, 113-0033, Japan}
\author{Nobunari Kashikawa}
\affiliation{Department of Astronomy, School of Science, The University of Tokyo, 7-3-1 Hongo, Bunkyo-ku, Tokyo, 113-0033, Japan}
\author{Jun Toshikawa}
\affiliation{Institute for Cosmic Ray Research, The University of Tokyo, 5-1-5 Kashiwa-no-Ha, Kashiwa, Chiba, 277-8582, Japan}
\affiliation{Department of Physics, University of Bath, Claverton Down, Bath, BA2 7AY, UK}
\author{Roderik Overzier}
\affiliation{Observat\'{o}rio Nacional, Rua Jos\'{e} Cristino, 77. CEP 20921-400, Sa\~{o} Crist\'{o} va\~{o}, Rio de Janeiro-RJ, Brazil}
\affiliation{Institute of Astronomy, Geophysics and Atmospheric Sciences, Department of Astronomy, University of S\~{a}o Paulo, Sao Paulo, SP 05508-090, Brazil}
\author{Mariko Kubo}
\affiliation{National Astronomical Observatory of Japan, Mitaka, Tokyo, 181-8588, Japan}
\author{Hisakazu Uchiyama}
\affiliation{National Astronomical Observatory of Japan, Mitaka, Tokyo, 181-8588, Japan}
\author[0000-0002-2725-302X]{Yongming Liang}
\affiliation{Department of Astronomical Science, The Graduate University for Advanced Studies, SOKENDAI, Mitaka, Tokyo, 181-8588, Japan} 
\affiliation{National Astronomical Observatory of Japan, Mitaka, Tokyo, 181-8588, Japan}
\author[0000-0003-2984-6803]{Masafusa Onoue}
\email{onoue@mpia-hd.mpg.de}
\affiliation{Max-Planck-Institut f\"ur Astronomie, K\"onigstuhl 17, D-69117 Heidelberg, Germany}
\author{Masayuki Tanaka}
\affiliation{National Astronomical Observatory of Japan, Mitaka, Tokyo, 181-8588, Japan}
\affiliation{Department of Astronomical Science, The Graduate University for Advanced Studies, SOKENDAI, Mitaka, Tokyo, 181-8588, Japan}
\author{Yutaka Komiyama}
\affiliation{National Astronomical Observatory of Japan, Mitaka, Tokyo, 181-8588, Japan}
\affiliation{Department of Astronomical Science, The Graduate University for Advanced Studies, SOKENDAI, Mitaka, Tokyo, 181-8588, Japan} 
\author{Chien-Hsiu Lee}
\affiliation{National Optical Astronomy Observatory 950 N Cherry Avenue, Tucson, AZ 85719, USA}
\author{Yen-Ting Lin}
\affiliation{Institute of Astronomy and Astrophysics, Academia Sinica, Taipei 10617, Taiwan}
\author{Murilo Marinello}
\affiliation{Laborat\'{o}rio Nacional de Astrof\`{i}sica - Rua dos Estados Unidos 154, Bairro das Na\c{c}\~{o}es, CEP 37504-364, Itajub\'{a}, MG, Brazil}
\author{Crystal L. Martin}
\affiliation{Department of Physics, University of California, Santa Barbara, CA 93106, USA}
\author{Takatoshi Shibuya}
\affiliation{Kitami Institute of Technology, 165 Koen-cho, Kitami, Hokkaido 090-8507, Japan}



\begin{abstract}
We report the rest-frame ultraviolet luminosity function of $g$-dropout galaxies in 177 protocluster candidates (PC UVLF) at $z\sim4$ selected in the Hyper Suprime-Cam Subaru Strategic Program. Comparing with the UVLF of field galaxies at the same redshift, we find that the PC UVLF shows a significant excess towards the bright-end. This excess can not be explained by the contribution of only active galactic nuclei, and we also find that this is more significant in higher dense regions. Assuming that all protocluster members are located on the star formation main sequence, the PC UVLF can be converted into a stellar mass function. Consequently, our protocluster members are inferred to have a 2.8 times more massive characteristic stellar mass than that of the field Lyman break galaxies at the same redshift. This study, for the first time, clearly shows that the enhancement in star formation or stellar mass in overdense regions can generally be seen as early as at $z\sim4$. We also estimate the star formation rate density (SFRD) in protocluster regions as $\simeq 6-20\%$ of the cosmic SFRD, based on the measured PC UVLF after correcting for the selection incompleteness in our protocluster sample. This high value suggests that protoclusters make a non-negligible contribution to the cosmic SFRD at $z\sim4$, as previously suggested by simulations. Our results suggest that protoclusters are essential components for the galaxy evolution at $z\sim4$.
\end{abstract}


\section{Introduction} \label{sec:intro}
\par Properties of galaxies are known to be correlated to their environments. Galaxies in local clusters tend to be early-type \citep[e.g,][]{Dressler80}, older \citep[e.g,][]{Thomas05}, and redder \citep[e.g,][]{Bamford09} than galaxies in blank field. However, it is still unclear when and how such environmental trends are shaped. Exploring environmental trends in the early universe, when such difference can emerge for the first time, is thus important for solving this long-standing question.
\par At higher redshifts ($z\geq2$), we have some overdense regions called protoclusters, which are defined as structures that will collapse into virialized objects with $M_{\rm halo}\geq10^{14}{\rm M_\odot}$ at $z\geq0$ \citep[see][for a comprehensive review]{Overzier16}. These structures are not yet virialized, unlike clusters, and most of them consist of star-forming galaxies instead of quiescent ones. Protoclusters have been found through a large variety of selection techniques. In terms of galaxies as tracers of the overdensity, some studies have used line-emitting galaxies such as H$\alpha$ emitters (HAEs) \citep[e.g.,][]{Hayashi:2012ht,2011MNRAS.415.2993H} and Ly$\alpha$ emitters \citep[e.g.,][]{2019ApJ...879...28H,2018NatAs...2..962J,Toshikawa12,2005ApJ...620L...1O,2002ApJ...569L..11V}, while others have focused on sub-millimeter galaxies (SMGs) \citep[e.g.,][]{2018Natur.556..469M}, or continuum detected ones such as photo-$z$ selected galaxies \citep[e.g.,][]{Chiang14} and Lyman break galaxies \citep[LBGs, e.g.,][]{Toshikawa16,2008ApJ...673..143O,Steidel98}. Also, several studies have used intergalactic medium (IGM) as tracers, such as by Ly$\alpha$ tomography \citep[e.g.,][]{2016ApJ...817..160L,2015MNRAS.453.4311S,2014ApJ...788...49L} or strong coherent Ly$\alpha$ absorption along the line of sight, so-called ``CoSLAs" \citep[e.g.,][]{2016ApJ...833..135C}.
\par Protocluster galaxies at $z\sim2$ have been shown to differ their properties compared to field galaxies at the same epoch. They tend to have enhancements of star formation rates (SFRs) \citep[e.g.,][]{2018MNRAS.473.1977S, 2013MNRAS.428.1551K}, with larger stellar mass \citep[above references and][]{2014MNRAS.440.3262C,2011MNRAS.415.2993H,2005ApJ...626...44S}. This suggests that the galaxy formation is earlier in protoclusters, as supported by several theoretical studies \citep{2017ApJ...844L..23C,2018MNRAS.474.4612L, 2015MNRAS.452.2528M}. Moreover, these theoretical studies suggest that these differences are already in place at even higher redshift. The examination of the galaxy population in protoclusters at higher redshifts is thus crucial for understanding effects of environment.
\par However, the SFR and the stellar mass of galaxies in overdense regions at $z\geq3$ have not yet been comprehensively assessed. There are several reasons for this. First, only $\sim 20$ protoclusters have been found to date at $z\geq3$ due to their extremely low number density \citep{Overzier16}, which is insufficient for a systematic study. Second, the target selection is highly heterogeneous: in addition to the variety of tracers of galaxies mentioned above, some studies focus on regions around quasars or radio galaxies \citep[e.g.,][]{Hayashi:2012ht,Venemans07}, while others focus on blank fields \citep[e.g.,][]{Toshikawa16, Chiang14}. Third, the precise estimation of the stellar mass and SFR through spectral energy distribution (SED) fitting requires the rest-frame optical data. At $z\geq4$, the rest-frame optical is shifted to the (near) infrared ($\lambda\geq2.0\ {\rm\mu m}$) in the observed frame, so observations become much more challenging. 
\par The rest-frame ultraviolet (rest-UV) luminosity function (UVLF) is an effective and practical tool for unraveling the properties of high redshift galaxies. The rest-UV light is generally emitted from short-lived massive stars and thus a good tracer of SFR \citep{1998ARA&A..36..189K}. UVLFs of field galaxies as a function of the cosmic time are the dominant diagnostic for understanding the history of cosmic star formation \citep[e.g.,][and see \citealt{2014ARA&A..52..415M} for a comprehensive review]{2015ApJ...803...34B,2012A&A...539A..31C,VanderBurg2010}. In addition, if we apply a relation between the stellar mass and SFR, so-called ``main sequence" \citep[e.g.,][and references therein.]{Song16, Speagle14}, UVLFs provide shapes of galaxy stellar mass functions (SMFs). Therefore, estimating a UVLF of protocluster galaxies at $z\geq3$ will provide us with an opportunity of revealing the general properties of galaxies in high-density regions. On the other hand, an accurate measurement of UVLFs of protoclusters requires a large number of protocluster samples, which has been the biggest obstacle.
\par Recently, we have conducted a new protocluster survey \citep[hereafter called T18]{Toshikawa18} from the photometric data of the Hyper Suprime-Cam (HSC) Subaru Strategic Program (HSC-SSP) \citep{Aihara18a}. Starting with a map of the overdensity of LBGs at $z\sim4$ (so-called $g$-dropout galaxies), defined as the difference in the local surface number density of galaxies from its average, we have found 179 protocluster candidates over an area of $121\ {\rm deg^2}$. Based on this sample, we have conducted several follow-up studies, investigating the relation between overdensity and bright QSOs \citep{2018PASJ...70S..32U}, and quasar pairs \citep{2018PASJ...70S..31O}, considering the brightest UV-selected galaxies in protoclusters as candidates of proto-brightest cluster galaxies \citep{2019ApJ...878...68I}, and using the stacked infrared (IR) properties of protoclusters to probe obscured star formation and active galactic nuclei (AGNs) \citep{2019ApJ...887..214K}. The systematic and homogeneous selection combined with the large size of our protocluster sample should also enable us to estimate the general UVLF of protocluster galaxies at $z\sim4$ for the first time. 
\par In this paper, we present the first measurement of the UVLF of galaxies in protoclusters at $z\sim4$. The remainder of this paper is organized as follows. We introduce our protocluster sample and their member galaxies in Section \ref{sec:2} and describe the procedure and results of the measurement of the UVLF in Section \ref{sec:3}. The SMF, the variety of UVLF, and the SFR density (SFRD) of their member galaxies inferred from the UVLF are estimated in Section \ref{sec:4}. Section \ref{sec:5} examines the validity of this result and discusses the implications for the galaxy formation in overdense regions. We summarize the paper in Section \ref{sec:6}. In this paper, we assume that cosmological parameters are $H_0= 70\ {\rm km\ s^{-1}\ Mpc^{-1}}$, $\Omega_{\rm m}=0.3$, and $\Omega_\Lambda=0.7$. We use the AB magnitude system.
\section{Data Summary, Sample Selection}\label{sec:2}
\par In this paper, we use protocluster candidates and the galaxy catalog constructed in T18. They draw overdensity maps of $g$-dropout galaxies from HSC-SSP S16A internal data release, which is a part of PDR1 \citep{Aihara18b}. Here, we briefly summarize the procedure for the selection of $g$-dropout galaxies and protocluster candidates.
\subsection{Galaxy Selection}\label{sec:galSel}
\par T18 use the HSC-SSP S16A internal data release for selecting $g$-dropout galaxies. HSC is the prime focus camera of the Subaru Telescope \citep{2018PASJ...70S...1M, Komiyama18}. 
\par The HSC-SSP survey is a wide and deep survey of over 300 nights by the HSC collaboration \citep{Aihara18a}. The target fields are divided into three layers (Wide, Deep, and UltraDeep), and five broad bands ($grizy$) and three narrow bands are used \citep[for more details on the HSC filter system, see][]{Kawanomoto18}. The Wide layer has a $5\sigma$ limiting magnitude of $i\sim26$ mag. HSC-SSP data is processed via {\tt hscpipe} \citep{Bosch18}, which is a modified version of the Legacy Survey of Space and Time software \citep{Juric15,Axelrod10, Ivezic08}. In the S16A data release, the total survey area of the Wide layer observed in all bands and reaching to the full depth is $178\ {\rm deg^2}$, and the average seeing is $0.56\arcsec$ in $i$ band and $0.65\arcsec-0.7\arcsec$ in other bands.
\par T18 construct a $g$-dropout galaxies sample from the $gri$ band photometry.  Only five regions in the Wide layer have enough depth (XMM-LSS, WIDE12H, GAMA15H, HECTOMAP, and VVDS) to construct a homogeneous map of the galaxy distribution. T18 impose color criteria (for $g-r$ and $r-i$) and a limiting magnitude cutoff ($5\sigma$ significance in the $i$ band and $3\sigma$ significance in the $r$ band), based on the {\tt Cmodel} magnitudes \citep{Bosch18}. Various flags are used to select objects with the clean photometry and not affected by cosmic rays and so on (For more detail, see T18).
\subsection{Protocluster Selection}\label{sec:PCsel}
\par T18 select protocluster candidates according to the peak value of the overdensity significance. The overdensity map of $g$-dropout galaxies is drawn from their surface number density through the fixed aperture method. This method distributes circular apertures on an every $1\arcmin$ grid and estimates the surface number density of galaxies from the number of galaxies inside the apertures. They define the aperture size of $1.8\arcmin$, which corresponds to $\sim 0.75$ physical Mpc at $z\sim 3.8$. This size is the smallest one expected for protoclusters of ``Fornax-type" clusters ($M_{\rm halo}\sim1-3\times10^{14}{\rm M_\odot}$ at $z\sim0$), as predicted by simulations \citep{2013ApJ...779..127C}. 
\par T18 only focuses on regions whose limiting $5\sigma$ magnitudes for $g,\ r,\ i$ band are deeper than 26.0, 25.5, and 25.5 mag, respectively, giving an effective survey area of 121 ${\rm deg^2}$. For drawing the overdensity map, T18 utilizes the $g$-dropout galaxies that are brighter than $25$ mag in $i$ band. T18 select 179 overdense regions whose peak overdensity significance is greater than $4\sigma$ as protocluster candidates, following \citet{Toshikawa16}. T18 evaluate that about $\geq76\%$ of such regions will evolve into halos with a mass greater than $10^{14} {\rm M_\odot}$ at $z\sim0$.
\par This large sample of protoclusters allows T18 to conduct an angular clustering analyses and estimate the mean dark matter halo mass as $\langle M_{\rm halo }\rangle=2.3_{-0.5}^{+0.5}\times10^{13}h^{-1}{\rm M_\odot}$. According to the extended Press-Schechter model, halos with such a large mass is indeed expected to evolve into those with $\langle M_{\rm halo}\rangle=4.1_{-0.7}^{+0.7}\times10^{14}h^{-1}{\rm M_\odot}$ at $z\sim0$.
\par We have to define the volume of protoclusters in order to measure the UVLF. It should be noted that these protocluster candidates and their members have the redshift uncertainty ($\delta z\sim1$) since this method is based on the dropout technique. We approximate the shape of protoclusters as cylinders. The cross-section of the cylinder is a circle with a radius of $1.8\arcmin$ corresponding to $0.75$ physical Mpc, which is the same size as the aperture in the overdensity map. The line-of-sight length is equivalent to the diameter of the cross-section. Therefore, we select protocluster member galaxies from galaxies that are located within a projected $<1.8\arcmin$ from the center of the overdensity peak. We consider a masked region in determining protocluster volumes. Note that we do not consider the particular morphology of each protocluster. For example, some protoclusters, particularly more massive ones, can be bigger \citep[e.g.,][]{2015MNRAS.452.2528M,2013ApJ...779..127C}. Some studies also argue that the shape of protoclusters can be described in the triaxial model \citep{2018MNRAS.474.4612L}. The radius for selecting member galaxies in the study is the minimum size of protoclusters predicted by the simulation \citep{2013ApJ...779..127C}, thereby our selected regions are expected to contain pure protocluster members, but we might miss some member galaxies that are located in the outermost regions of protoclusters. As we discuss in Section \ref{sec:5-2}, our results for the shape of UVLF do not significantly change even if we change the radius of the cross-section and the depth. 
\section{Rest-UV Luminosity Function Measurement}\label{sec:3} 
\subsection{Formulation of Luminosity Function}\label{sec:F_LF}
\par We estimate the UV absolute magnitude, which is the absolute magnitude at $1500\ {\rm \AA}$ in the rest-frame from the apparent magnitude. As mentioned in Section \ref{sec:PCsel}, our protocluster galaxies have a significant redshift uncertainty since they are selected from $g$-dropout galaxies. Therefore, we fix $\bar{z}=3.8$ as the typical redshift. We convert the $i$ band magnitude $(m_i)$ by using the following equation;
\begin{equation}
    M_{\text{UV}}=m_i+2.5\log{(1+\bar{z})}-5\log{(\frac{d_L(\bar{z})}{10\ {\rm pc}})}+(m_{1500(1+\lambda)}-m_i)
    \label{eq:MUV}
\end{equation}
\par Here, $d_L(\bar{z})$ is the luminosity distance at $z=\bar{z}$ in the unit of pc. We assume that the $g$-dropout galaxies' SED at rest-UV is flat in $f_\nu$, which leads to a $k$-correction factor ($m_{1500(1+\lambda)}-m_i$) of zero, following \citet{Ono2018}.
\par We measure only the projected number density from the photometric data; therefore, our protocluster galaxy sample has some possible contaminants. One is fore/background $g$-dropout galaxies outside the protocluster regions, hereafter called ``field galaxies". The effective redshift range of $g$-dropout galaxies is significantly larger than the protocluster's transverse size, so we must subtract the contribution of field galaxies from the measured surface number density in protocluster regions. The number density of field galaxies can be approximated by the UVLF of field galaxies (field UVLF) since the volume fraction of protocluster is small compared to the total survey volume. In addition to field galaxies, $g$-dropout galaxies themselves may inevitably have some contaminants such as stars and low-redshift galaxies due to the color selection uncertainties, which should be removed from the sample. These objects can be assumed to be homogeneously distributed if we combine all protoclusters, which are separated on the whole sky; therefore, their contamination rate should be the same both inside and outside of the protocluster regions. This implies that the subtraction of the field UVLF without the contamination correction provides a clean estimate of the number density of protocluster galaxies.
\par One possible contamination source that is hard to assume to distribute homogeneously is low-$z$ galaxy clusters at $0.3<z<0.6$, where Balmer breaks are hardly distinguishable from Lyman break at $z\sim4$. \citet{2018PASJ...70S..20O} construct a galaxy cluster sample at $0.1<z<1.1$ from $232\ {\rm deg^2}$ HSC-SSP data. They find 620 clusters at $0.3<z<0.6$, implying their surface number density as $2.67\ {\rm deg^{-2}}$. The possibility that our protoclusters are overlapped with galaxy clusters at $0.3<z<0.6$ within $1.8 \arcmin$ (i.e., protocluster size) is only 0.59\%. Therefore, we conclude that all contamination is negligible to estimate the the UVLF of protocluster galaxies (PC UVLF).
\par We correct the effective volume of $g$-dropout galaxies to the protocluster effective volume by a factor $F$ defined as;
\begin{equation}
    F(M_{\text{UV}})=\frac{\langle C(M_{\text{UV}},z)\frac{dV(z)}{dz}\delta z\rangle }{V_{\text{eff}}(M_{\text{UV}})}
    \label{eq:F}
\end{equation}
\par Here, $C(M_{\text{UV}},z)$ is the completeness function of the $g$-dropout selection estimated in Section \ref{sec:CF}. $\delta z$ is the redshift interval that corresponds to the depth of the cylinder volume of protoclusters (see Section \ref{sec:PCsel}). $dV(z)/dz$ is the differential comoving volume. The $V_{\text{eff}}(M_{\text{UV}})$ is the effective volume for $g$-dropout galaxies in $1.8\arcmin$ aperture, which is defined as follows \citep[e.g.,][]{1999astro.ph..5116H};
\begin{eqnarray}
    V_{\text{eff}}(M_{\text{UV}}) &=& \int C(M_{\text{UV}},z)\frac{dV(z)}{dz}dz 
\end{eqnarray}
\par The numerator of $F(M_{\text{UV}})$ corresponds to the effective volume of a protocluster, whose shape is defined in Section \ref{sec:PCsel}. Therefore, $F(M_{\text{UV}})$ is the ratio of the effective volume of the protoclusters and the effective volume of the redshift range of the entire $g$-dropout selection. Since we do not know the exact redshifts of each system, we use the average numerator weighted by the redshift selection function (i.e., the completeness function).     
\par Then, the PC UVLF is described as follows,
\begin{equation}
    \Phi_{\text{PC}}(M_{\text{UV}}) = \frac{1}{F(M_{\text{UV}})}(\frac{n_{\text{obs,PC}}(M_{\text{UV}})}{V_{\text{eff}}(M_{\text{UV}})}-\Phi_{\text{field}}(M_{\text{UV}}))\label{eq:PCLF} 
\end{equation}
where $n_{\text{obs,PC}}(M_{\text{UV}})$ is the observed number of $g$-dropout galaxies in protocluster regions defined in Section \ref{sec:PCsel} in each magnitude bin. $ \Phi_{\text{field}}(M_{\text{UV}})$ is the field UVLF without the contamination correction (see Section \ref{sec:FieldLF}). In order to determine $\Phi_{\text{PC}}(M_{\text{UV}})$, we estimate the completeness function of $g$-dropout galaxies $C(M_{\text{UV}},z)$ and the field UVLF without contamination treatment in the following section.
\subsection{Completeness Estimation}\label{sec:CF}
\par As in the previous studies of UVLFs of field LBGs \citep[e.g.,][]{Ono2018, VanderBurg2010,2006ApJ...653..988Y}, we insert mock galaxies into actual images and estimate a completeness function. This is derived as a function of the redshift and the magnitude.
\par  Mock galaxies are inserted into the coadd images of the $g,r,i$ band images of HSC-SSP products. We generate mock images through the {\tt Balrog}\footnote{\url{https://github.com/emhuff/Balrog}} \citep{2016MNRAS.457..786S}, which inserts mock galaxies with the help of the {\tt galsim}\footnote{\url{https://github.com/GalSim-developers/GalSim}} \citep{2015A&C....10..121R} followed by their detection and measurement through {\tt SourceExtractor}. However, the HSC-SSP source catalog is constructed based on {\tt hscpipe}; therefore, we detect and measure the photometry of the mock galaxies through {\tt hscpipe}, instead. We use {\tt hscpipe} version 4, which is the same software used for the HSC-SSP S16A data release.
\par We assume that the surface brightness profile follows the S\'ersic profile \citep{Sersic63} with a fixed S\'ersic index of $1.5$ for mock galaxies. In addition, the effective size distribution is assumed to be consistent with that of \citet{2015ApJS..219...15S}. The real profile of mock galaxies are considered with the point spread function (PSF) of that field by convolving it taken from {\tt PSFEx}\footnote{\url{https://www.astromatic.net/software/psfex}} \citep{2011ASPC..442..435B}. The SED of mock galaxies are generated using the {\tt CIGALE}\footnote{\url{https://cigale.lam.fr/}} \citep{2019A&A...622A.103B}. We assume a constant star formation and use the single stellar population models of \citet{Bruzual03}. We adopt the Salpeter initial mass function (IMF) \citep{1955ApJ...121..161S} with an age of $100$ Myr and metallicity of $Z/Z_\odot=0.2$. The dust extinction follows \citet{Calzetti00} with $E(B-V)=0.0-0.4$ mag. The IGM absorption is accounted for according to \citet{2006MNRAS.365..807M}. We change their redshift from $3.0$ to $5.0$ with interval of $\delta z\sim 0.1$.
\par Due to the slight differences in depths among the five fields, we estimate the completeness function for each field. We select one region called {\tt tract}, with an area of ${\rm 2.3\ deg^2}$, for each field to execute the procedure. The number of inserted galaxies is about $35$ per ${\rm arcmin^2}$. From the detected catalogs, we select mock $g$-dropout galaxies by the same criteria as used in T18, including color, magnitude, and flags selection.
\par
\begin{figure*}
    \centering
    \includegraphics[width=160mm]{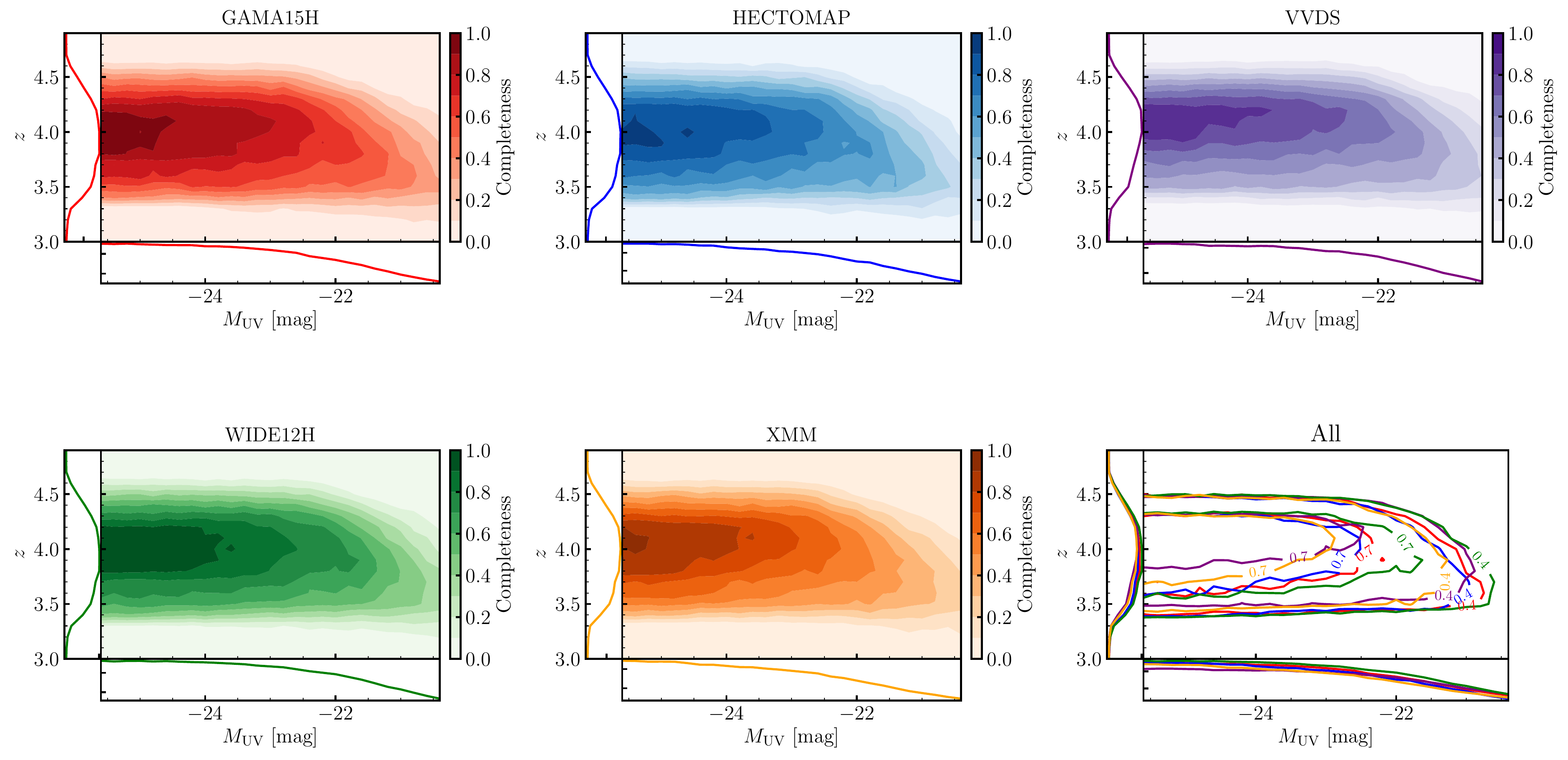}
    \caption{Completeness functions for each target field. The bottom right panel shows their comparison. Colors match those shown in each panel of single field. Values written in the contour in the right bottom panel represent the completeness.}
    \label{fig:CF}
\end{figure*}

 For each field, we calculate the completeness as the number ratio of selected mock $g$-dropout galaxies to all inserted objects in each magnitude and redshift bin. Figure \ref{fig:CF} shows the completeness function of each field, demonstrating that five fields have almost the same completeness. 
\subsection{Luminosity Function of Field Galaxies Without Contamination Treatment}\label{sec:FieldLF}
\par From the completeness function and Equation \ref{eq:fieldLF} below, we estimate the field UVLF without contamination treatment.
\begin{equation}
    \Phi_{\text{field}}(M_{\text{UV}}) = \frac{n_{\text{obs,field}}(M_{\text{UV}})}{V_{\text{eff}}(M_{\text{UV}})} 
    \label{eq:fieldLF}
\end{equation}
\par Here, $n_{\text{obs, field}}(M)$ is the observed number of field galaxies and contaminants of $M_{\text{UV}}=M$. Before deriving $n_{\text{obs, field}}(M)$, we remove all known low-$z$ galaxies, stars, or QSOs from the available spectroscopic survey archives, such as SDSS DR12 \citep{2015ApJS..219...12A}, HectoMAP cluster survey \citep{2018ApJ...856..172S}, and VIPERS DR1 \citep{2014A&A...562A..23G}. The majority of matched objects are galaxies at $0.3<z<0.6$ and QSOs at the same redshift distribution of $g$-dropout galaxies. Only two QSOs, which overlap with the protoclsuter region, are removed from the sample \citep{2018PASJ...70S..32U}.  
\par We compare the input total magnitude and the measured $2.0\arcsec$ aperture magnitude of mock galaxies used in Section \ref{sec:CF}, and find that $2.0\arcsec$ aperture magnitude has a $+0.08$ mag offset on average from the input magnitude. Therefore we apply a 0.08 mag aperture correction to our measured $2.0\arcsec$ magnitudes to derive the total magnitudes. We confirm that the derived total magnitudes are consistent with measured aperture magnitudes with the larger apertures, such as $3.0\arcsec,\ 4.0 \arcsec$. We also correct the galactic extinction by using the extinction map from \citet{Schlegel98}.
\par
\begin{figure}
    \centering
    \includegraphics[width=8.5cm]{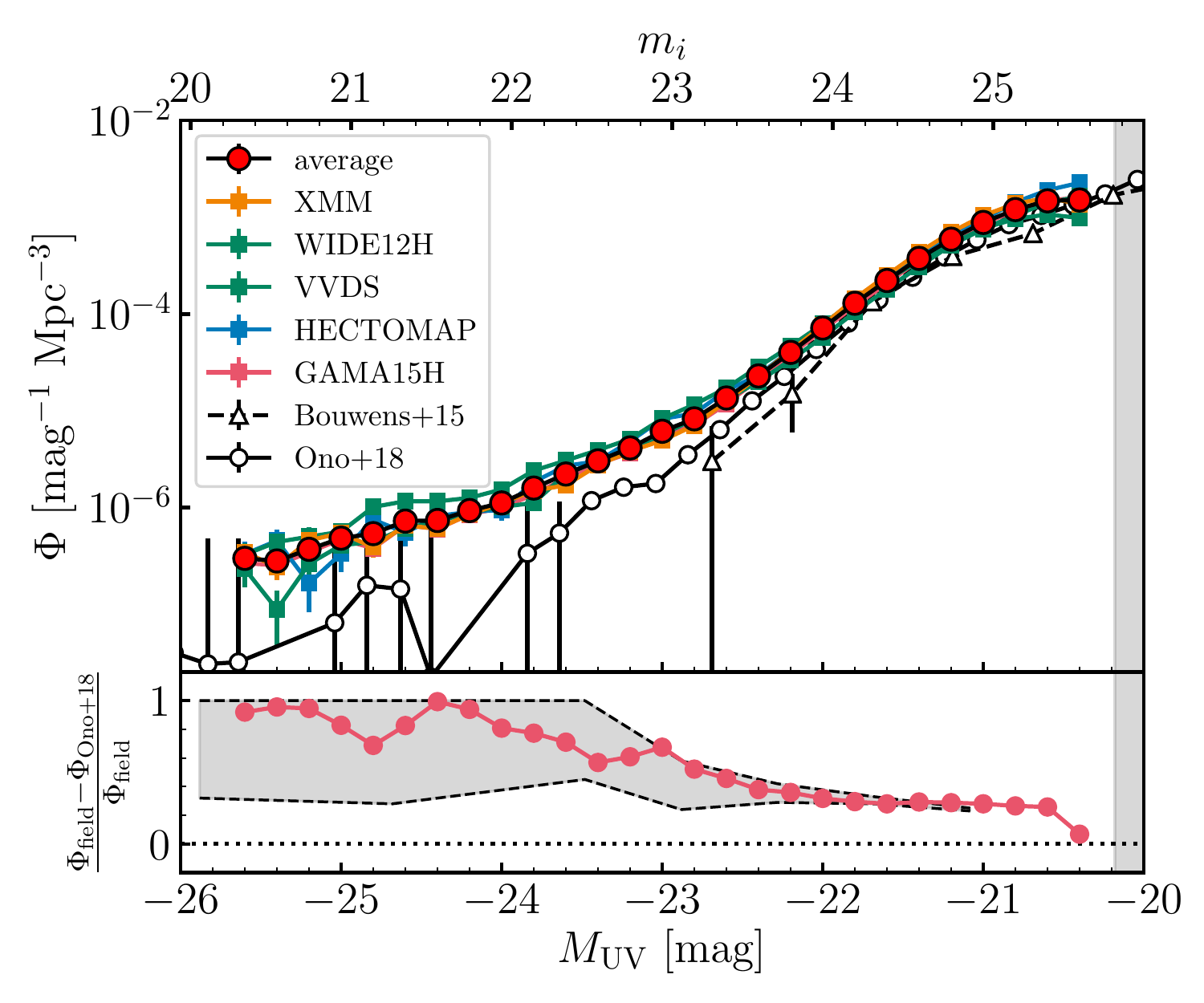}
    \caption{Top panel: The field UVLF at $z\sim4$ for each field (squares) and their average (red circles). Black open circles and triangles are UVLFs in the literature \citep{Ono2018,2015ApJ...803...34B}. Note that we do not correct for contaminants. Bottom panel: The red circles show the difference between the average of this work UVLFs, which is not corrected for contaminants, and the UVLF from \citet{Ono2018} normalized by this work's UVLF. Gray shaded region shows the contamination rate of $g$-dropout galaxies estimated in \citet{Ono2018}. The red circles correspond to a contamination fraction of our $g$-dropout galaxies, which is in good agreement with that in \citet{Ono2018}.}
    \label{fig:LF}
\end{figure}

 This UVLF is not necessarily the same as the field UVLF derived in previous studies \citep[e.g.,][]{Ono2018,2015ApJ...803...34B} since our function includes contaminants, as seen in Figure \ref{fig:LF}. We derive the field UVLFs for each field.
\par The bottom panel of Figure \ref{fig:LF} shows the difference between the average of the UVLFs and that of \citet{Ono2018} normalized by this UVLF. Since the UVLF of \citet{Ono2018} exclude contaminants, this represents the expected fraction of contaminants among our $g$-dropout galaxies. We can find that this ratio is consistent with that in \citet{Ono2018}, overplotted in the bottom panel of Figure \ref{fig:LF}. We conclude that our completeness function is consistent with previous studies. Hereafter, we will use these UVLFs and the completeness function to estimate the PC UVLF.
\subsection{Protocluster Luminosity Function}\label{sec:LF_PC}
\par Here, we estimate the PC UVLF according to Equation \ref{eq:PCLF}. Two protoclusters are excluded since they are located in low-quality regions with quite shallow limiting magnitudes ($m\sim25.6$ mag for $5\sigma$ $i$-band limiting magnitude); thus 177 protocluster regions are used for estimating the PC UVLF. Since the completeness function and $\Phi_{\text{field}}(M_{\text{UV}})$ have been determined for each field, the PC UVLF is also estimated for each field separately, and we take the average weighted by the total area for each field as our final PC UVLF. We note that all PC UVLF for each field are overall consistent within the uncertainty.  
\par
\begin{figure*}
    \centering
    \includegraphics{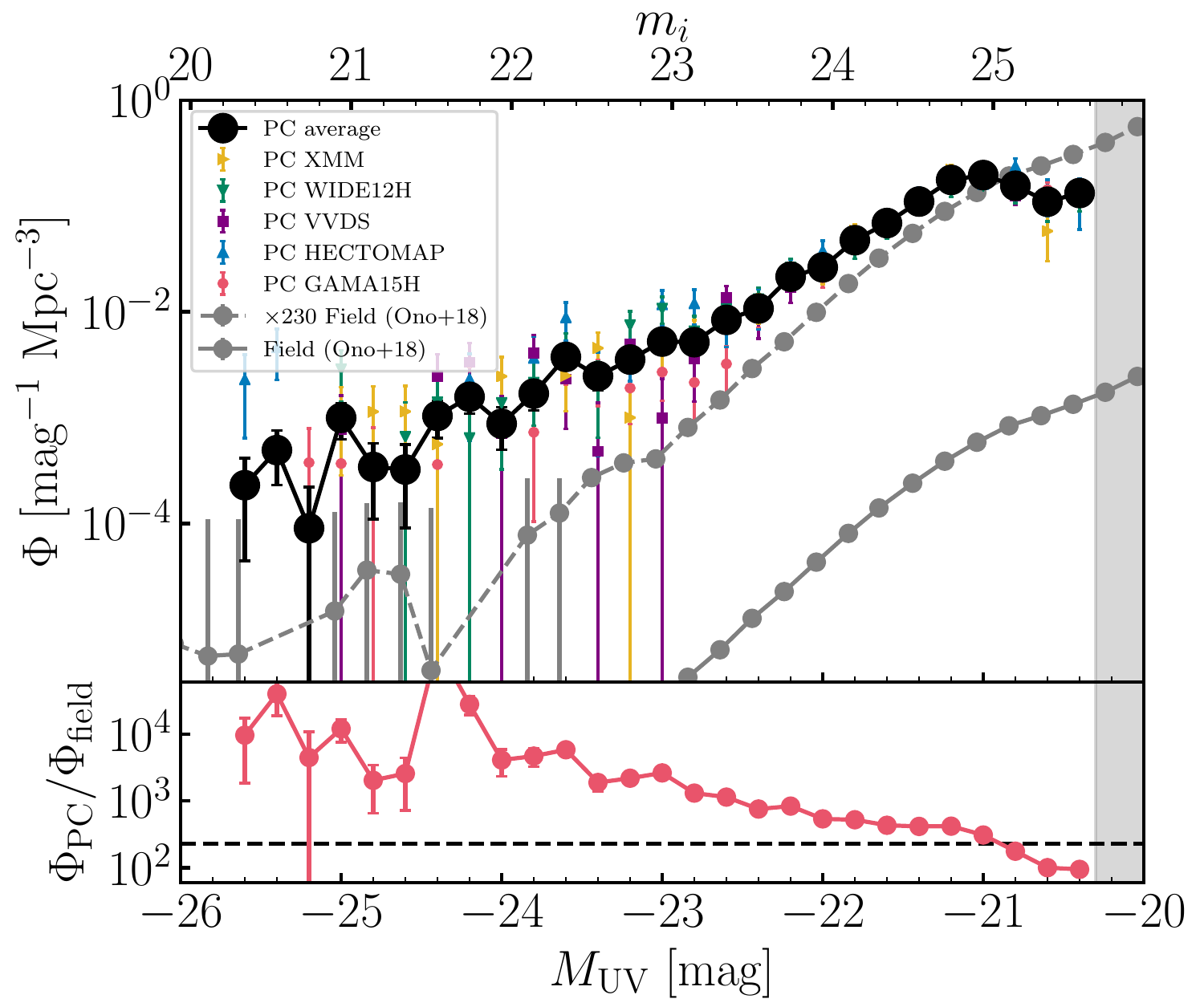}
    \caption{The luminosity function of galaxies in protocluster candidates at $z\sim4$. The color-coded markers represent the PC UVLF for each survey field. The black circles show the average of all fields. For reference, we show the field UVLF of \citet{Ono2018} (gray solid line with circles) and shifted upward to match the PC UVLF (gray dashed line with circles). The bottom panel shows the ratio of the PC UVLF and the field UVLF (red circles). The black dashed line shows the value of the ratio of the sum of each UVLF. For both panels, the magnitude range that is fainter than the depth is shaded in gray.}
    \label{fig:LF_PC}
\end{figure*}

 We show the average PC UVLF of the HSC-SSP protocluster candidates in Figure \ref{fig:LF_PC}. Our PC UVLF has apparent discrepancies with the field UVLF in the literature \citep[e.g.,][]{Ono2018}. First, the amplitude is much higher than the field UVLF, with the integrated value of the PC UVLF at $M_{\text{UV}}\leq-20.3$ is about 230 times higher than that of the field UVLF of \citet{Ono2018}. Second, its shape is remarkably different from the field UVLF. The amplitude-matched field UVLF is also shown in the top panel of Figure \ref{fig:LF_PC} for reference, and compared with that, the PC UVLF has a significant excess towards the bright-end ($M_{\text{UV}}\leq-20.8$). The trend can also be seen on the bottom panel of Figure \ref{fig:LF_PC}, which shows the ratio of the PC and the field UVLF. We see that the excess gets larger towards the brighter bin. If the shapes are identical between them, this ratio should stay constant at any brightness.
\par Since the number density of galaxies decreases towards the bright-end, so the photometric error of each galaxy might enhance the amplitude of the bright-end of UVLF, which is known as ``Eddington Bias" \citep{1913MNRAS..73..359E}. We estimate the effect of this bias by convolving the error distribution of magnitude to the field UVLF of \citet{Ono2018}. The detail of this analysis is described in Appendix \ref{sec:App1}. We confirm that the Eddington bias is not significant to generate the shape of our PC UVLF.
\par Since contributions from low-$z$ contaminants, which distribute homogeneously, are statistically subtracted from the sample as mentioned in Section \ref{sec:F_LF}, the bright-end excess is not due to low-$z$ galaxy contaminants. Also, the PC UVLF may depend on $F(M_{\text{UV}})$, which is the ratio of the effective volume of protoclusters and $g$-dropout galaxies. We confirm that the bright-end excess of PC UVLF does not change even if we fix $F (M_{\rm UV})=1$, as seen in Appendix \ref{sec:App2}.
\par The rest-UV luminosity of galaxies represents their SFR. Therefore, this result indicates that overdense regions at $z\sim4$ have not only a high SFRD caused by the excess of the number of galaxies, but also a higher fraction of galaxies with high SFR compared to those in the blank field. This trend is also seen in some protoclusters at lower redshifts. For example, \citet{2018MNRAS.473.1977S} estimate the SFR of HAEs in a protocluster at $z=2.5$, and they also find that HAEs in the densest regions tend to have a higher SFR than those in the outskirts. \citet{2013MNRAS.428.1551K} report a similar trend from HAEs in protoclusters at $z\sim2$. This paper, for the first time, shows that the enhancement of star formation of UV-bright galaxies in overdense regions can already be seen as early as from $z\sim4$. We have to note that some bright ($M_{\rm UV}<-23.0$) LBGs can be AGNs, whose UV emission cannot be a proxy of SFR of their host galaxies \citep[e.g.,][]{2019arXiv191201626A,Ono2018}. We discuss a possible contribution from AGN in Section \ref{sec:5-3}.
\subsection{Function Fitting}\label{sec:3-5}
\par To compare the shape of the PC UVLF with the field UVLF more quantitatively, we fit the Schechter function \citep{1976ApJ...203..557S}, which is defined as follows;
\begin{equation}
    \phi(L)dL=\phi^*\left(\frac{L}{L^*}\right)^\alpha \exp{\left(-\frac{L}{L^*}\right)}d\left(\frac{L}{L^*}\right)
\end{equation}
where $\alpha$ is the faint-end slope, $L^*$ is the characteristic luminosity, and $\phi^*$ is the overall normalization. This function can be also expressed as a function of the absolute magnitude $M_{\text{UV}}$, 
\begin{eqnarray}
\Phi(M_{\text{UV}})=\frac{\ln{10}}{2.5}\phi^*10^{-0.4(M_{\text{UV}}-M_{\text{UV}}^*)(\alpha+1)} \nonumber \\
     \times\exp(-10^{-0.4(M_{\rm UV}-M_{\rm UV}^*)})
\end{eqnarray}
\par
\begin{figure}
    \centering
    \includegraphics[width=8.5cm]{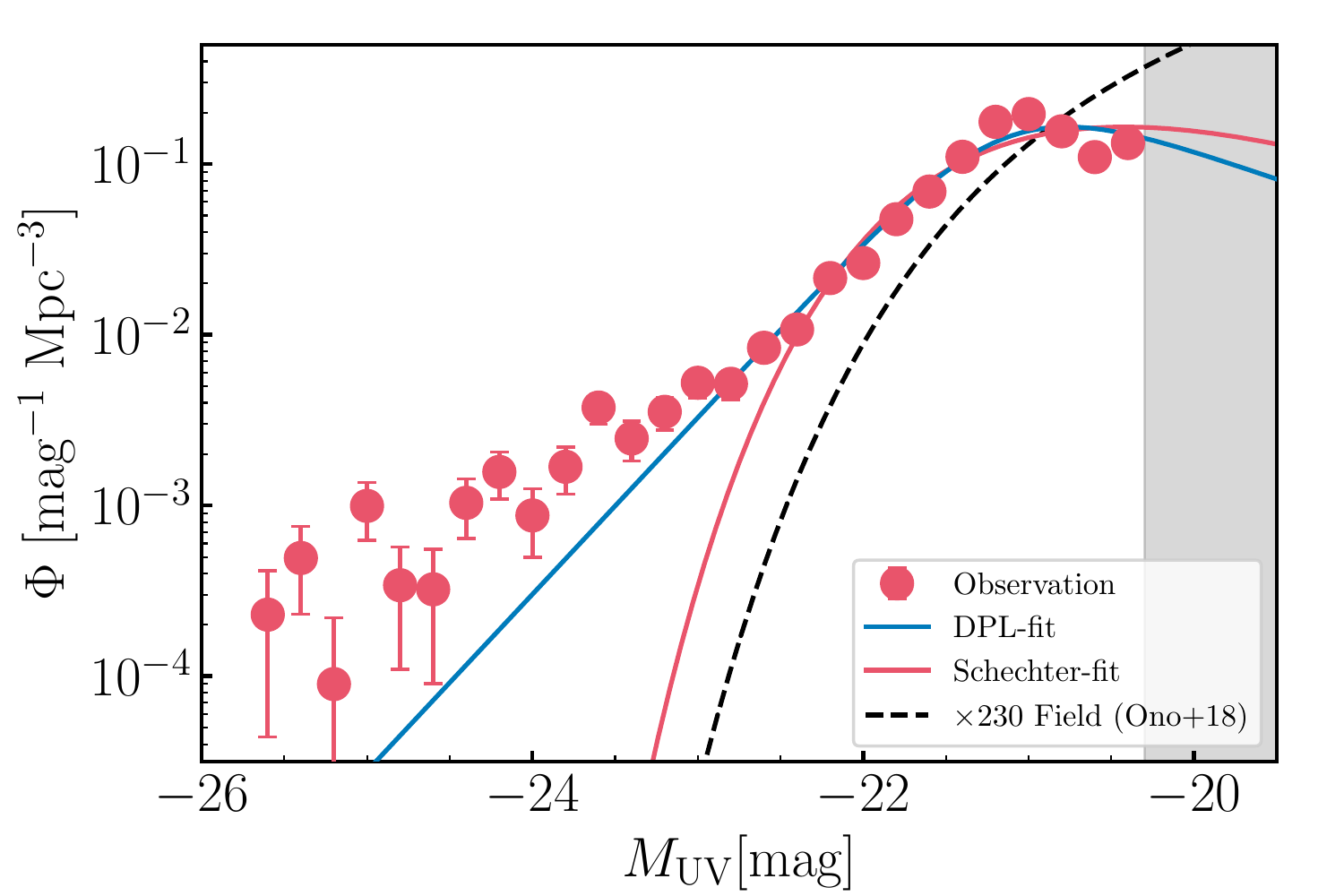}
    \caption{The result of the fitting of the Schechter/DPL function to PC UVLF. Circles show the derived PC UVLF. The red (blue) lines represent the best-fit of the Schechter (DPL) function. As a reference, the black dashed line is the best-fit Schechter function of the field UVLF in \citet{Ono2018}. Same as Figure \ref{fig:LF_PC}, the gray shade represents the magnitude range which is not discussed.}
    \label{fig:LF_model}
\end{figure}

\begin{figure*}
    \centering
    \includegraphics[width=15cm]{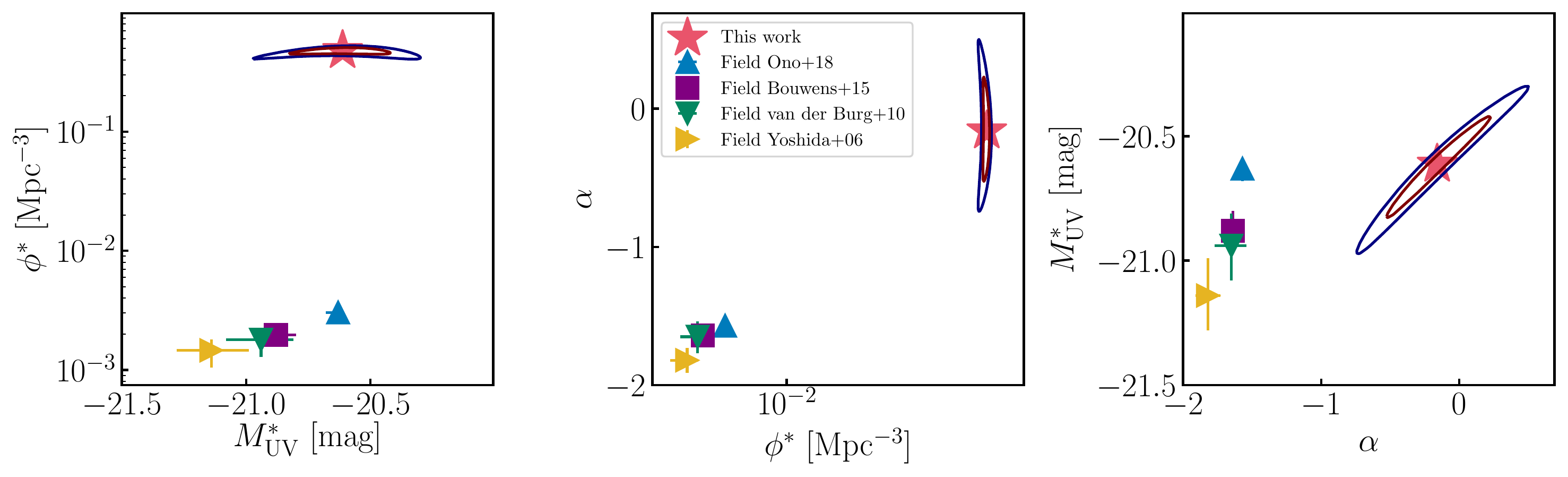}
    \caption{The comparison of best-fit parameters of our PC UVLF with those of the field UVLFs from the literature. Red stars represent this work, and blue, purple, green and yellow markers represent \citet{Ono2018}, \citet{2015ApJ...803...34B}, \citet{VanderBurg2010}, and \citet{2006ApJ...653..988Y}, respectively. Red and Blue contours represent the $68.3\%$, and $95.5\%$ confidence levels of the best-fit parameters of our PC UVLF, respectively.}
    \label{fig:LFparam}
\end{figure*}

 We fit the Schechter function in terms of absolute magnitude to the PC UVLF using the $\chi^2$ minimization method. We show the best-fit Schechter function in Figure \ref{fig:LF_model} and the parameters in Table \ref{tab:2}. Compared to the best-fit parameters of the field UVLF in previous studies \citep{Ono2018,2015ApJ...803...34B, VanderBurg2010,2006ApJ...653..988Y}, our PC UVLF has a less steep faint-end slope, as shown in Figure \ref{fig:LFparam}. Our best-fit $M^*_{\text{UV}}$ is consistent with that of the field UVLFs at the $68/95\%$ confidence level. This implies that the PC UVLF has a different shape compared to the field UVLF, although the discrepancy between our PC UVLF and the best-fit Schechter function is large, particularly at the bright-end ($M_{\text{UV}}<-23$).
\par The large reduced $\chi^2$ shown in Table \ref{tab:2} implies this failure of fitting at the bright-end. This can be because the PC UVLF does not seem to have a clear exponential decrease at the bright-end. Therefore, we try to fit another functional form. Recent UVLF studies of field galaxies at higher redshift ($z\geq4$) have suggested that the galaxy UVLF can be well described by a double power-law (DPL) function \citep[e.g.,][]{Bowler:2019tb,2015MNRAS.452.1817B, Ono2018}. The DPL function is defined as follows;
\begin{equation}
    \phi(L)dL=\phi^*\left[\left(\frac{L}{L^*}\right)^{-\alpha} +\left(\frac{L}{L^*}\right)^{-\beta}\right]^{-1}\frac{dL}{L*}
    \label{eq:DPL}
\end{equation}
\par ,where $\beta$ represents the power-law slope at the bright-end ($M_{\text{UV}}<M^*_{\text{UV}}$). We fit this function in terms of absolute magnitude, also. We fix the faint-end slope $\alpha$ to be the same as that of the best-fit Schecter function. We also show the best-fit DPL function in Figure \ref{fig:LF_model}, and their parameters in Table \ref{tab:2}. The DPL function fits better than the Schechter function, even though the best-fit DPL function still has some deviation from the observed PC UVLF at $M_{\text{UV}} < 23$. 
\par The excess from the best-fit Schechter/DPL function of UVLFs of field galaxies is often explained by AGNs. \citet{Ono2018} claim that the gap of UVLFs of field galaxies from their best-fit Schechter function at $z\sim4-7$ is explained by the contribution of AGN UVLFs at the same redshift. Also, \citet{2016ApJ...823...20K} construct the Ly$\alpha$ luminosity function of LAEs at $z=2.2$ and argue that the gap at the brightest-end from its best-fit is due to AGNs. We discuss the possible contribution from AGNs in Section \ref{sec:5-3} and do not reject the possibility of the gap in both best-fit results due to AGNs. However, we can not conclude which functions represent the galaxy UVLF more precisely. Therefore, we use both fitting functions in the following sections.
\begin{table}[]
    \centering
    \begin{tabular}{ccccc}
    \hline
    $M^*_{\rm UV}$  & $\phi^*$&$\alpha$&$\beta$&$\chi^2_\nu$ \\
    (${\rm mag}$) & (${\rm \ Mpc^{-3}}$) &&& \\
    \hline
    \multicolumn{5}{c}{Schechter function}\\
    $-20.61_{-0.14}^{+0.12}$&$0.48_{-0.02}^{+0.02}$&$-0.16_{-0.25}^{+0.25}$&-&11.2\\
    \hline
    \multicolumn{5}{c}{Double power-law function}\\
    $-21.13_{-0.04}^{+0.04}$ & $0.31_{-0.01}^{+0.01}$&$(-0.16)$&$-3.59_{-0.11}^{+0.08}$&$5.5$\\
    \hline
    \end{tabular}
    \caption{The best-fit parameters and the reduced $\chi^2$ of the Schechter and DPL functions fitted to the PC UVLF. We fix the faint-end slope in the case of the DPL to the best-fit value in the case of the Schechter function.}
    \label{tab:2}
\end{table}
\section{implications from the protocluster galaxy luminosity function}\label{sec:4}
\subsection{Stellar Mass Function}\label{sec:4-2}
\par We estimate the SMF based on the measured PC UVLF, assuming that all protocluster $g$-dropout galaxies are located on the star formation main-sequence of field galaxies at the same redshift. We utilize the main-sequence estimated by \citet{Song16}, which determine the main-sequence by applying SED-fitting analysis to field photo-$z$ selected galaxies from \citet{2015ApJ...810...71F}. We assume the main sequence is equivalent between protoclusters and the field, which is supported by observational studies \citep[e.g.,][]{Long:2020vn,2019ApJ...871...83S,2013MNRAS.428.1551K} and a theoretical study \citep[e.g.,][]{Lovell:2020wz}, while some studies report a large contribution from star burst galaxies in protoclusters \citep[e.g.,][]{2018Natur.556..469M}, leading to the possibility of different main sequence from that of field galaxies. 
\par
\begin{figure}
    \centering
    \includegraphics[width=8.5cm]{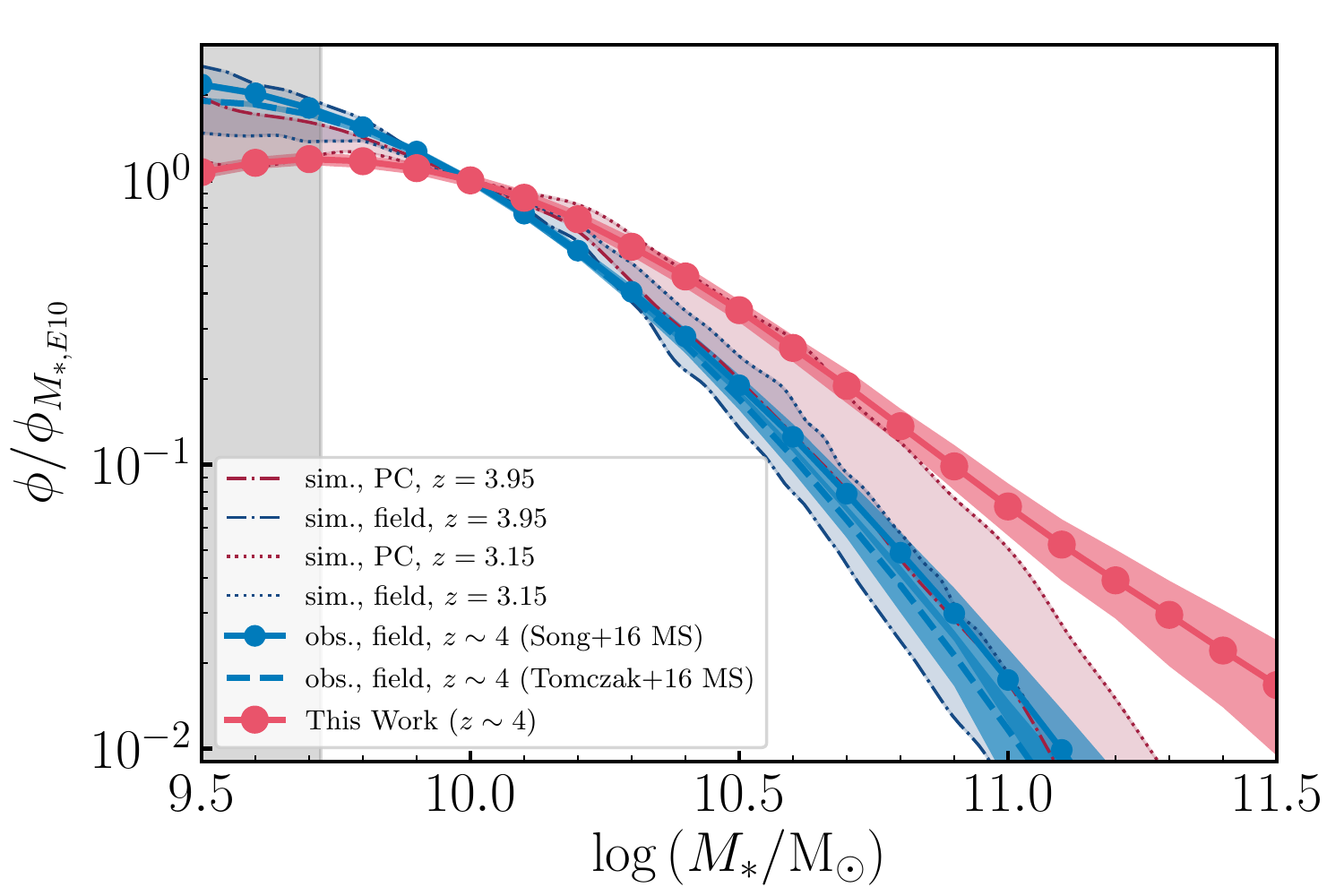}
    \caption{The comparison between our SMFs and SMFs predicted in \citet{2018MNRAS.474.4612L}. Red and blue circles are SMFs of protocluster and field galaxies estimated in this study. Their $1\sigma$ uncertainty are shown in shaded regions with each color. Blue dashed line is the SMF of field gaalxies from \citet{2016ApJ...817..118T}. Red and blue dash-dotted (dotted) lines are predicted SMFs of galaxies in protocluster and those in the field at z=3.95 (z=3.10), respectively. We normalize SMFs at $\log{(M_*/{\rm M_\odot})}=10.0$.}
    \label{fig:Ms_func_model}
\end{figure}

 We use the ``constant-scatter galaxy SMF" method, which is conducted in some previous studies \citep[e.g.,][]{Song16}. First, $M_{\text{UV}}$ is randomly assigned. Its probability distribution for each $M_{\text{UV}}$ is approximated by the PC UVLF, in which Gaussian random errors for each bin are assigned, whose $1\sigma$ is equivalent to that of the observed PC UVLF. The $M_{\text{UV}}$ is converted into the stellar mass $M_*$ according to the $M_*$-$M_{\text{UV}}$ relation of \citet{Song16} with a constant scatter of $0.4$ dex, and finally, the stellar mass distribution is obtained. This procedure is repeated for 1000 times, and the SMF of protocluster galaxies (PC SMF) is obtained by taking their average. The uncertainty of the SMF is taken from the variation among 1000 results. The SMF of field galaxies (field SMF) is also estimated from the field UVLF of \citet{Ono2018} in the same manner. We find that the estimation of SMFs has only a negligible change within the uncertainty when we use the main sequence of \citet{2016ApJ...817..118T}, which has a flatter massive end ($\log{(M_*/{\rm M_\odot})}>10.5$), compared to the main sequence of \citet{Song16} as shown in Figure \ref{fig:Ms_func_model}.
\par Figure \ref{fig:Ms_func_model} shows our SMF estimate. We normalize them to fix the value at $\log{(M_*/{\rm M_\odot})}=10$ for the easy comparison. The gray shaded region ($\log{(M_*/{\rm M_\odot})}<9.72$) in Figure \ref{fig:Ms_func_model} shows the incomplete mass range due to the limiting magnitude ($M_{\text{UV}}>-20.3$). We hereafter discuss the SMF in the stellar mass range of  $\log{(M_*/{\rm M_\odot})}>9.72$. The PC SMF shows a clear excess from that of field galaxies towards the massive end, suggesting that protoclusters contain a relatively high fraction of massive galaxies compared to the field. Here, we mention three notes. First, this SMF only includes $g$-dropout galaxies, which are typically star-forming, and we do not consider quiescent galaxies. Recent studies report the existence of massive quiescent galaxies even at $z\sim4$ in the blank field \citep[e.g.,][]{2020ApJ...889...93V,2019ApJ...885L..34T}, but the fraction of them are expected to be small ($<5\%$) according to field SMFs \citep[e.g.,][]{2017A&A...605A..70D}, though the value in overdense environments has uncertainty. Therefore, we ignore the effect of quiescent galaxies. Second, the bend of the PC UVLF around $M_{\rm UV}<-23$ is not seen in PC SMF. This is because the SMF is estimated from the main-sequence with the constant scatter, which is so called ``Eddington Bias". Third, the most massive-end ($\log{(M_*/{\rm M_\odot})}>11.15$) is dominated by objects with $M_{\text{UV}}\leq-23$. As we mention in Section \ref{sec:5-2}, objects in such magnitude range can be AGNs; therefore, values of the SMF in this mass range can have uncertainty.
\par
\begin{figure}
    \centering
    \includegraphics[width=8.5cm]{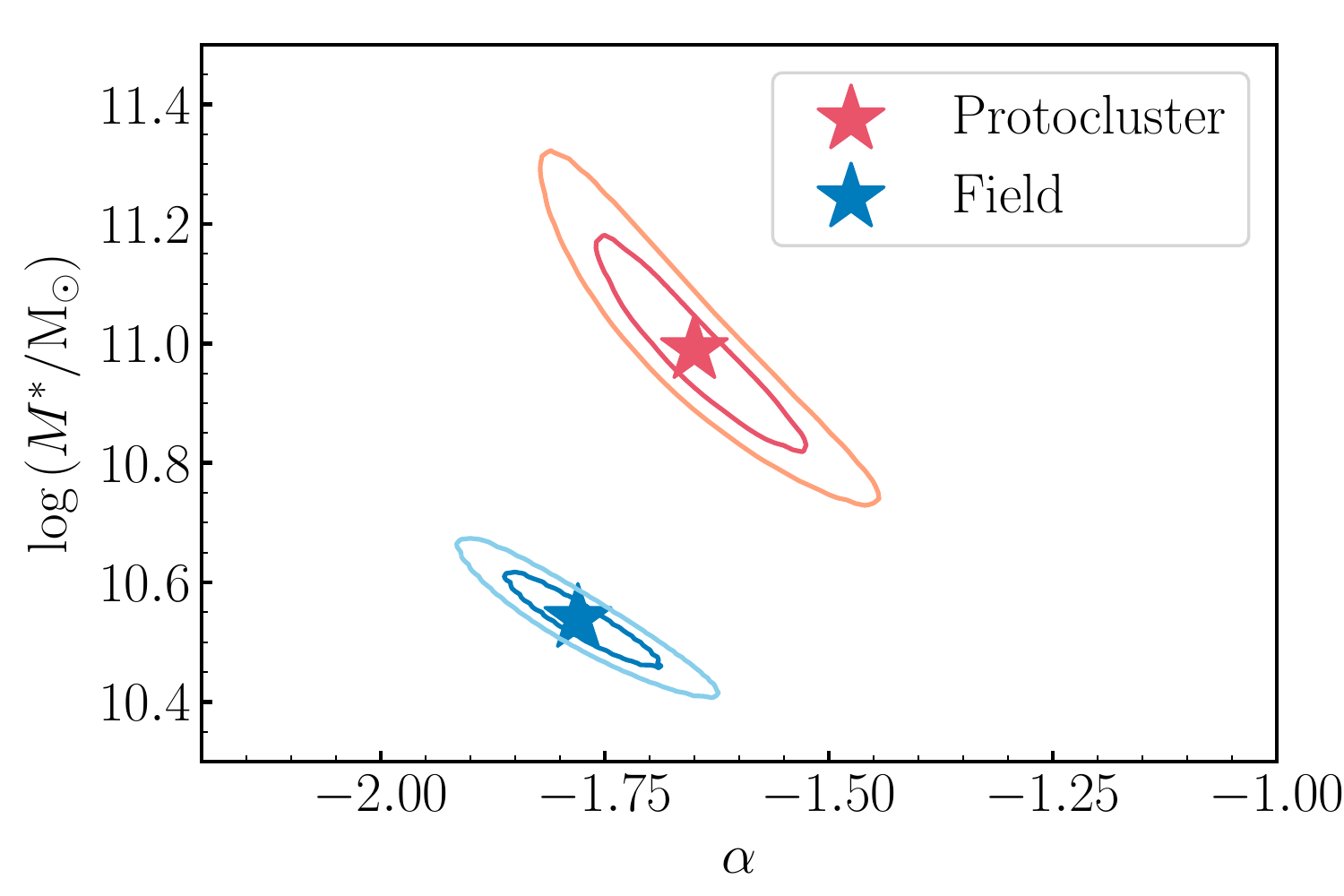}
    \caption{The best-fit parameters of the Schechter function to the PC SMF (Red star) and the field SMF (Blue star). The contours represent their $68/95\%$ confidence interval.}
    \label{fig:MF_param}
\end{figure}

 We fit the Schechter function to the measured PC SMF as well as to the field SMF at $z\sim4$. We can see that the PC SMF has a higher characteristic stellar mass and faint-end slope than the field SMF as seen in Figure \ref{fig:MF_param}. Protocluster galaxies have about 2.8 times higher characteristic stellar mass than field galaxies. This also supports the result that protocluster galaxies are more massive than field galaxies.
\par  The difference of the PC SMF and the field SMF is also seen in simulations at $z\sim4$ \citep{2018MNRAS.474.4612L,2015MNRAS.452.2528M}. In Figure \ref{fig:Ms_func_model}, we compare our PC SMF and the field SMF with those predicted in \citet{2018MNRAS.474.4612L}. \citet{2018MNRAS.474.4612L} use the semi-analytical model (SAM) from \citet{2015MNRAS.451.2663H} and trace the evolutionary track of halos with $M_{200}/{\rm M_\odot}>10^{14}$ at $z\sim0$ to higher redshift. $M_{200}$ is the mass within $r<r_{200}$, where the density is 200 times the critical density. We use their predicted SMFs constructed from galaxies with ${\rm SFR}>5\ {\rm M_\odot}\ {\rm yr}^{-1}$ at $z=3.10\ \text{and} \ 3.95$. The average redshift of our protocluster sample is between redshifts of these predicted SMFs. Our SMF is found to be almost consistent with the theoretical predictions and located between the predicted SMF at $z=3.95$ and that at $z=3.10$. Though the PC SMF has higher amplitude than the theoretical prediction at the most massive-end ($\log{(M_*/{\rm M_\odot})}>11.15$), this can be explained by the contribution of AGNs mentioned above.
\par
\begin{figure}
    \centering
    \includegraphics[width=8.5cm]{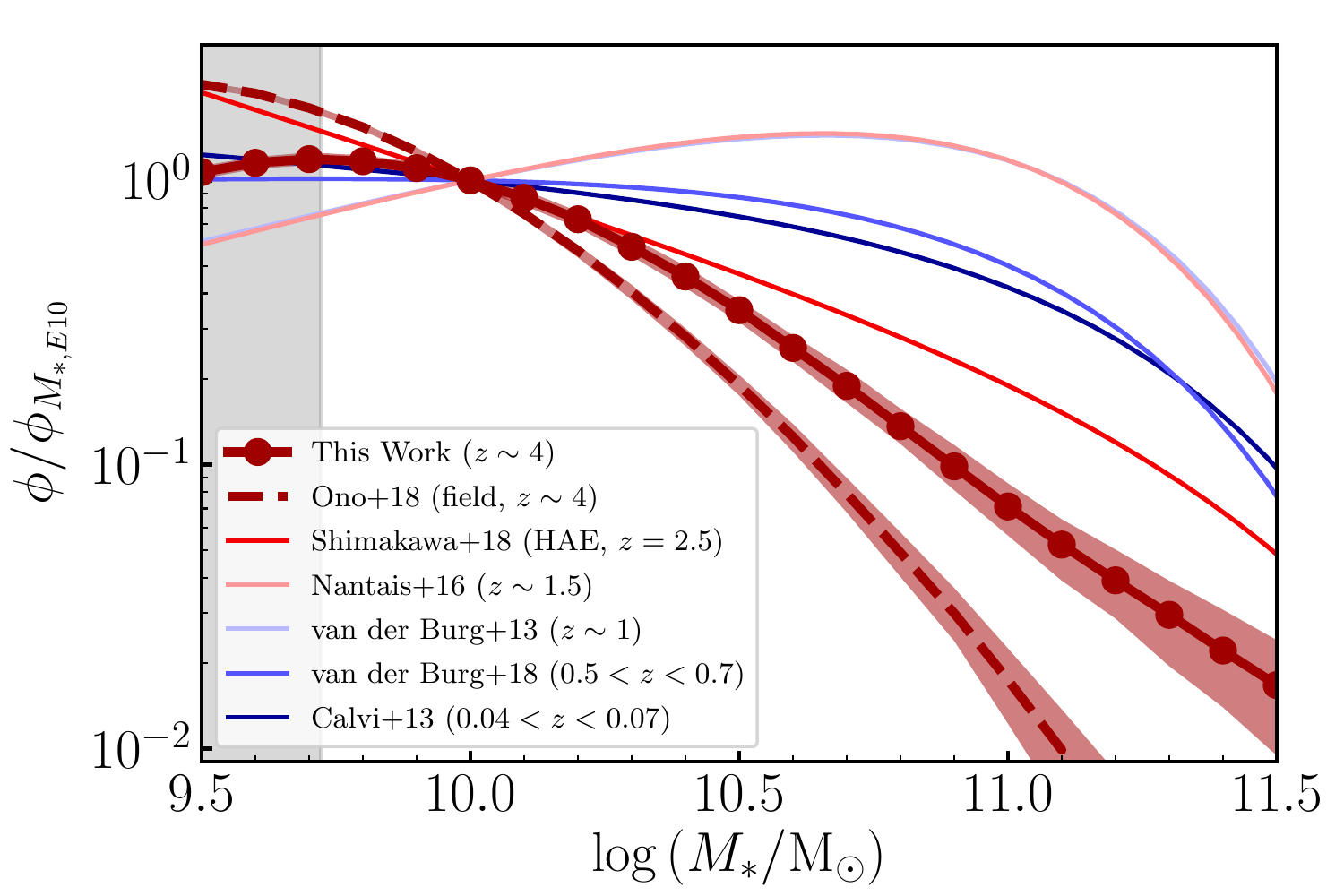}
    \caption{SMFs of (proto)cluster galaxies at different redshift. Their amplitudes are normalized  at $\log{(M_*/{\rm M_\odot})}=10.0$. A red line with circles show that of protocluster galaxies at $z\sim 4$ (this work). For reference, the SMF of field galaxies estimated from the field UVLF of \citet{Ono2018} is plotted in a red dashed line. Their shaded region shows the $1\sigma$ uncertainty of them. \added{Other red lines show SMFs of $z\geq1.5$ (proto)clusters from other studies \citep{2018MNRAS.473.1977S,2016A&A...592A.161N}, and blue lines show SMFs of $z\leq1$ clusters \citep{vanderBurg:2013gl,2018A&A...618A.140V,2013MNRAS.432.3141C}}.}
    \label{fig:Ms_func}
\end{figure}

 We compare our PC SMF with those of (proto)cluster galaxies at lower redshifts. \citet{2018MNRAS.473.1977S} estimate a SMF of HAEs in a protocluster called USS1558-003 at $z\sim2.5$. \citet{2016A&A...592A.161N} focus on four galaxy clusters at $z\sim1.5$ from the Spitzer Adaptation of the Red-sequence Cluster Survey (SpARCS) \citep{2009ApJ...698.1934M,2009ApJ...698.1943W,2010ApJ...711.1185D}. \citet{vanderBurg:2013gl} present a SMF of galaxies of ten rich clusters in the Gemini Cluster Astrophysics Spectroscopic Survey (GCLASS) at $0.86<z<1.34$. The SMF of galaxies in 21 clusters detected with the Plank satellite at $0.5<z<0.7$ is also presented in \citet{2018A&A...618A.140V}.  \citet{2013MNRAS.432.3141C} estimate a SMF of cluster galaxies from the WIde-field Nearby Galaxy-cluster Survey (WINGS) at $0.04\leq z\leq 0.07$ \citep{2006A&A...445..805F}, and compare with that of field galaxies at the same redshift. Figure \ref{fig:Ms_func} shows our PC SMF with other SMFs and the field SMF. Same as in Figure \ref{fig:Ms_func_model}, we normalize the amplitude of all SMFs at $\log{(M_*/{\rm M_\odot})}=10$. This is because the definition of the (proto)clusters' volume depends on studies, leading to the difficulty of the amplitude comparison. Therefore, we only focus on the shape difference of these SMFs. We also convert their assumed IMF to Salpeter IMF, which is used in \citet{Song16}.
\par We can see that there is a dearth of massive galaxies in the SMF of our protoclusters at $z\sim4$ compared to those at lower-$z$. This suggests that our protoclusters at $z\sim4$ are still in the process of mass growth. Particularly, from $z\sim4$ (HSC-SSP protoclusters) to $z\sim1$ \citep{vanderBurg:2013gl}, SMFs shows a monotonic growth at the massive end. At $z\sim0-1$, the ratio of SMFs at massive-end to that at low mass-end decreases towards lower redshift. This may be due to the significant contribution of less massive infalling galaxies. We discuss it in more detail in Section \ref{sec:5-4}. 
\par We note that these SMFs are based on galaxy clusters selected by different methods. They might be at different stages of the evolution of clusters \citep{2020ApJ...888...89T}, which may make it difficult to compare them with each other. Moreover, protocluster sample of this study and \citet{2018MNRAS.473.1977S} only focus on star-forming galaxies, while others contain quiescent galaxies. The fraction of quiescent galaxies at $z>2$ is known to be smaller than that at lower redshift, so we ignore the effects of this difference. Also as mentioned in Section \ref{sec:galSel}, our protocluster candidates are overdense regions expected to evolve into clusters with $\langle M_{\rm halo}\rangle= 4.1^{+0.7}_{-0.7}\times10^{14}h^{-1}{\rm M_\odot}$ at $z\sim0$. The majority of clusters from WINGS is as massive as $M_{200}\sim(1-10)\times10^{14}{\rm M_\odot}$ \citep{2017A&A...607A..81B}, which is same mass range as the expected halo mass of our protoclusters. On the other hand, the cluster halo mass of other studies is  $M_{200}\sim3\times10^{14}{\rm M_\odot}$ for SpARCS \citep{2012MNRAS.427..550L} and GCLASS \citep{vanderBurg:2013gl}, and $M_{200}\sim(3-13)\times10^{14}{\rm M_\odot}$ in \citet{2018A&A...618A.140V}. These clusters are already as massive as WINGS clusters, even at $z\sim1$, so they may grow more by $z\sim0$, leading them to have difficulty for comparing with WINGS clusters and our sample. In addition, the halo mass of USS1558-003 is not estimated; therefore, it is still under debate whether HSC-SSP protoclusters at $z\sim4$ are progenitors of protoclusters such as USS1558-003.  
\subsection{The Diversity of Protocluster Luminosity Functions}\label{sec:4-3}
\par Our protocluster sample has some variation in terms of overdensity. As shown in Figure 1 of \citet{2018PASJ...70S..32U}, the overdensity of protoclusters ranges from $4\sigma$ to $9.5\sigma$, and overdensity and descendant halo mass are broadly positively correlated \citep{Toshikawa16}. Here, we make subsamples of protoclusters according to the overdensity and construct UVLFs for each subsample.
\par
\begin{figure}
    \centering
    \includegraphics[width=8.5cm]{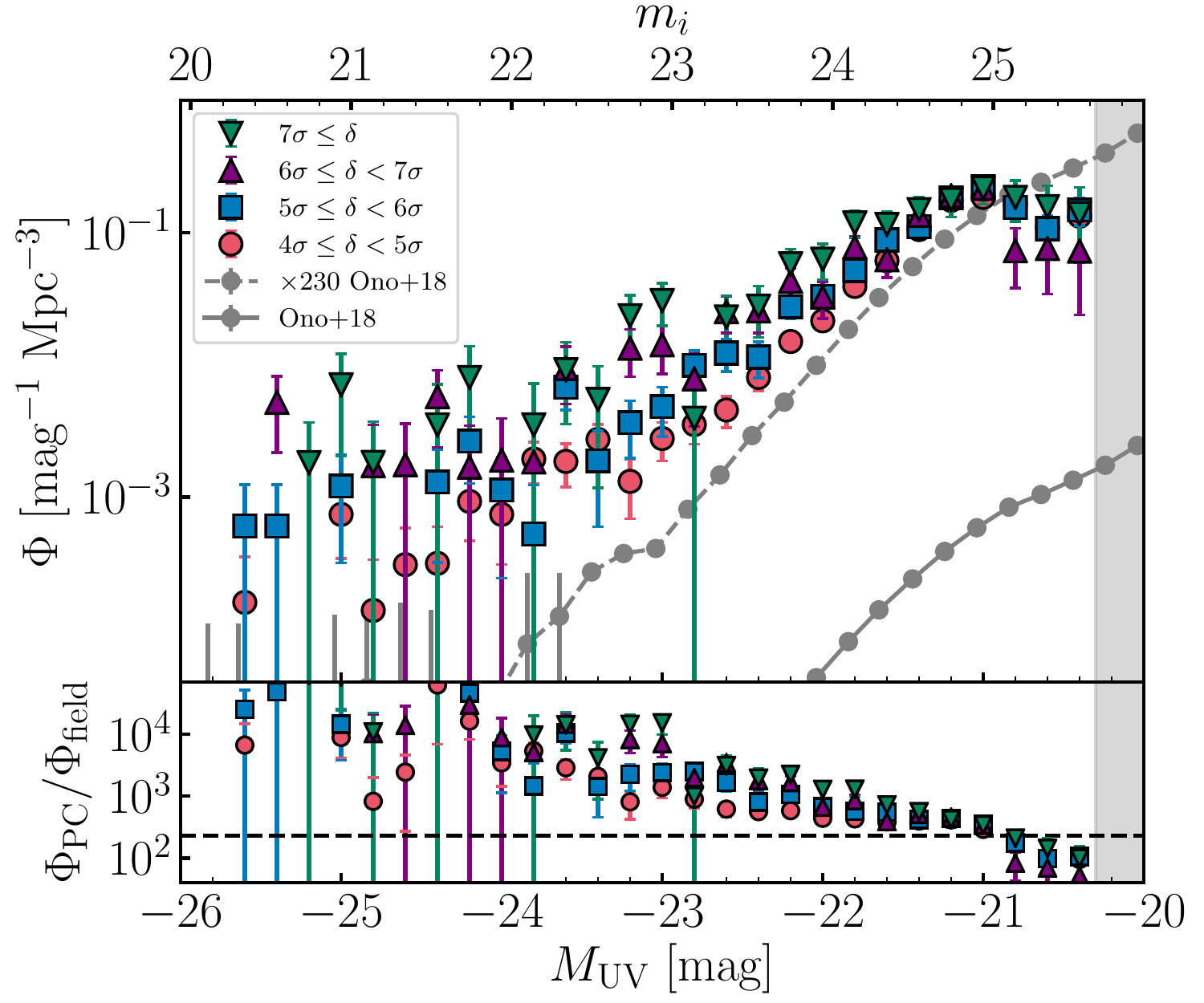}
    \caption{The UVLFs of members of protoclusters grouped according to their overdensities. Red, blue, purple, and green markers show those whose host protoclusters' overdensity are $4\sigma\leq\delta<5\sigma$, $5\sigma\leq\delta<6\sigma$, $6\sigma\leq\delta<7\sigma$, $7\sigma\leq\delta$, respectively. The gray lines are same as in Figure \ref{fig:LF_PC}.}
    \label{fig:PCLF_os}
\end{figure}

 We divide protocluster samples into four groups according to their overdensity $\delta$; 1). $4\sigma\leq\delta<5\sigma$, 2). $5\sigma\leq\delta<6\sigma$, 3). $6\sigma\leq\delta<7\sigma$, 4). $7\sigma\leq\delta$. The numbers of protoclusters in each subgroup are 120, 37, 13, and 7, respectively. In Figure \ref{fig:PCLF_os}, we show the PC UVLF for each subsample. The amplitude of the faint-end ($M_{\text{UV}}>-21.2$) is almost the same among subsamples, while the bright-end  ($M_{\text{UV}}<-21.2$) depends on the overdensity of protoclusters. More overdense protoclusters tend to have a higher bright-end amplitude compared to less massive protoclusters. These protoclusters can be more spatially extended, which could cause such a dependency on overdensity; however, we find that this is unlikely as discussed in Section \ref{sec:5-2}.
\par
\begin{figure}
    \centering
    \includegraphics[width=8.5cm]{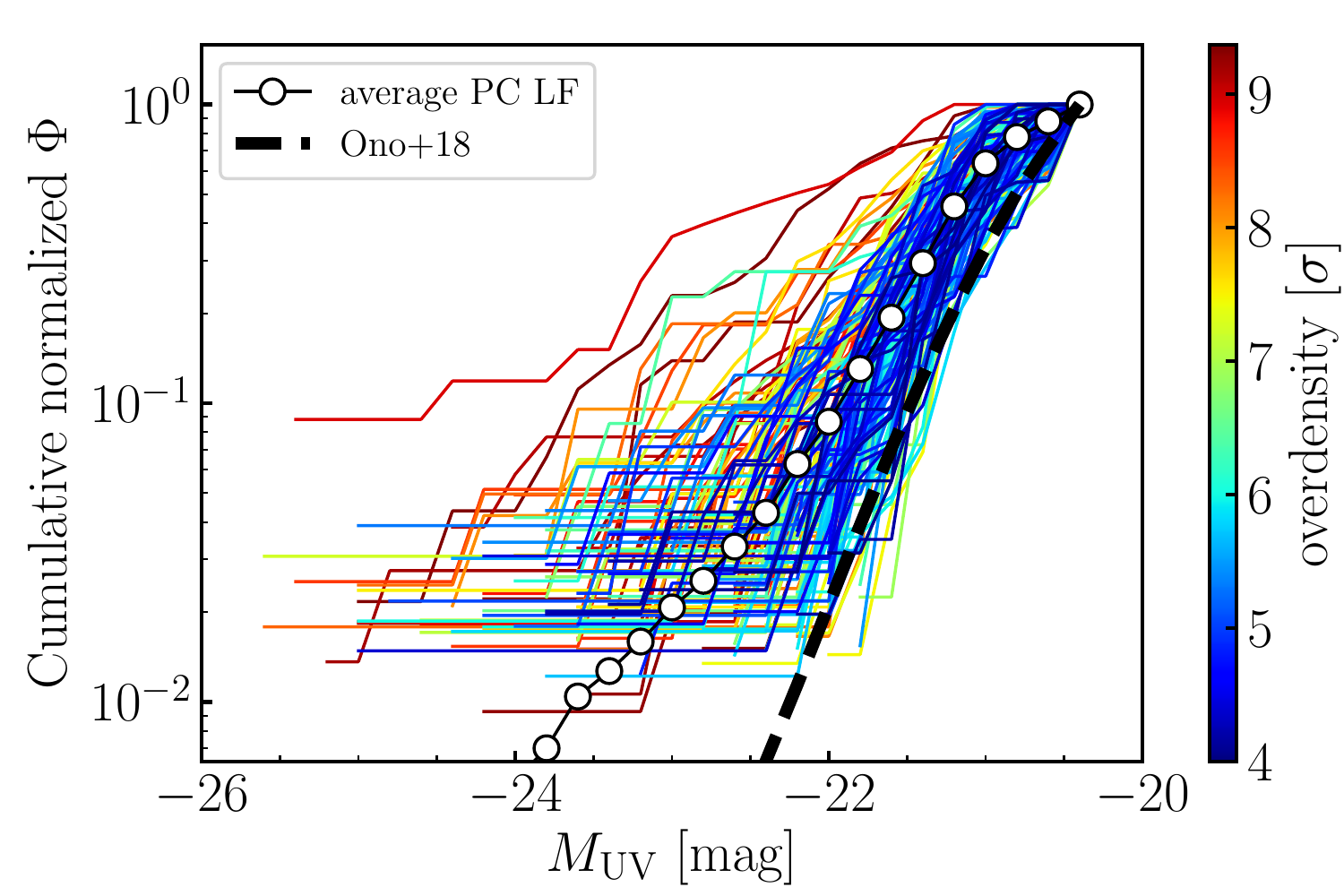}
    \caption{The cumulative UVLFs of galaxies in each protocluster candidates. Colors represent their overdensity significance. For reference, the average cumulative luminosity function of protocluster galaxies  (black open circles) and the cumulative luminosity function of field galaxies (dashed line) are also plotted.}
    \label{fig:LF_PCind}
\end{figure}

 The dependency of the bright-end excess on overdensity can be seen even for each protocluster separately. Figure \ref{fig:LF_PCind} shows the cumulative UVLF of galaxies in each protocluster. The bright-end amplitude of more overdense protoclusters tends to be higher than those of less massive protoclusters, suggesting that protoclusters with higher overdensity significance have brighter objects. More interestingly, almost all of protoclusters at $z\sim4$ have this excess at the bright-end compared to those of field galaxies, although the variation is seen even if we focus on only protoclusters with the same overdensity. Therefore, we conclude that the bright-end excess is ubiquitously seen for protoclusters at $z\sim4$. 
\par In \citet{2019ApJ...878...68I}, we investigate the significantly UV-brightest galaxies (proto-BCGs) in this protocluster sample. We find that galaxies in protoclusters containing proto-BCGs are brighter than other protocluster galaxies. This can be due to the overdensity dependence of the bright-end excess since the average overdensity of protocluster containing proto-BCGs is slightly higher ($(5.068\pm0.149)\sigma$) than that of all protoclusters ($(4.767\pm0.069)\sigma$). To reach the cause of the bright-end excess, we divide a subgroup, which is made in this subsection, into two according to whether protoclusters contain proto-BCGs. At a fixed overdensity, the UVLF of members of protoclusters containing proto-BCGs has the same bright-end amplitude to those of protocluster not containing proto-BCGs. Brighter galaxies of protoclusters containing proto-BCGs are thus due to their higher overdensity.
\subsection{Star Formation Rate Density}\label{sec:4-1}
\par We estimate the SFRD of protocluster galaxies, based on a combination of the PC UVLF and the far IR (FIR) luminosity density. The PC UVLF is approximated by the best-fit Schechter/DPL function. Parameter spaces with a $68\%$ confidence level estimated in Section \ref{sec:3-5} are employed for the PC UVLF.
\par We first estimate the UV luminosity density $\rho_{\text{UV}}$ from the PC UVLF as $\rho_{\text{UV}}=\int_{L_{\text{faint}}}^{L_{\text{bright}}} L_{\text{UV}}\phi(L_{\text{UV}})dL_{\text{UV}}$. We set $L_{\text{faint}}=2.7\times10^{27} {\rm erg\ s^{-1}\ Hz^{-1}}$, corresponding to $M_{\text{UV}}=-17$ mag, which is the same as applied in \citet{2015ApJ...803...34B}, and $L_{\text{bright}}=1.1\times10^{31} {\rm erg\ s^{-1}\ Hz^{-1}}$, corresponding to $M_{\text{UV}}=-26$ mag.
\par The FIR ($8-1000\mu m$) luminosity density $\rho_{\text{FIR}}$ is estimated as $\rho_{\text{FIR}}=\int_{L_{\text{faint}}}^{L_{\text{bright}}} L_{\text{FIR}}\phi(L_{\text{UV}})dL_{\text{UV}}$ with the use of the IRX-$\beta$-$M_*$ relation of $z\sim3$ LBGs \citep{2019A&A...630A.153A}. 
The $\beta-M_{\text{UV}}$ relation is known to exist even in protocluster galaxies at $z\sim4$ \citep{2008ApJ...673..143O}. The $\beta$ distribution is determined by using the conversion equation from $i-y$ color to $\beta$ in \citet{2012ApJ...754...83B}. We linearly fit the median value of $\beta$ distribution in each 0.2 mag magnitude bin of $M_{\text{UV}}\leq-20.3$. We use its best-fit parameters with their $1\sigma$ error for $\beta-M_{\text{UV}}$ relation.  We also estimate the $\beta-M_{\text{UV}}$ relation of our field galaxies in the same manner and compare it with the literature in Appendix \ref{sec:App3}. Our estimation is consistent with literature within the uncertainty, suggesting that our measurement and the sample selection is robust. The stellar mass $M_*$ is estimated from the UV absolute magnitude in the same method in Section \ref{sec:4-2} with the correction of IMF from Salpeter IMF to that of what \citet{2019A&A...630A.153A} use \citep{2003ApJ...586L.133C} by dividing stellar mass by 1.74. From the $\beta-M_{\text{UV}}$ relation and the estimated stellar mass, $L_{\text{UV}}$ is converted into $L_{\text{FIR}}$. 
\par We derive average $\rho_{\text{UV}}$ and $\rho_{\text{FIR}}$ weighted by the likelihood obtained in the fitting. We employ their minimum and maximum value to estimate the error by varying the parameters of the UVLF/$\beta-M_{\text{UV}}$ relation in the range of their 16th and 84th percentiles, respectively. As a result, we estimate the UV/FIR luminosity density of HSC-SSP protocluster galaxies as $\rho_{\text{UV}} = 3.46^{+0.35}_{-0.29}\times10^{28}\ ( 3.53^{+0.17}_{-0.16}\times10^{28})\ {\rm erg\ s^{-1}\ Hz^{-1}\ {\rm Mpc^{-3}}}$, and $\rho_{\text{FIR}} = 1.7^{+0.9}_{-0.9}\times 10^{11}\ (2.5^{+1.8}_{-1.0} \times10^{11})\ L_\odot\ {\rm Mpc^{-3}}$ in the case of the Schechter (DPL) function, respectively.
\par \citet{2019ApJ...887..214K} conduct stacking analysis of FIR images taken from Planck, AKARI, IRAS, and Herschel at the position of HSC-SSP protoclusters, which is the same sample in this study. Based on their best-fit of the SED model composed of star, dust and AGN flux components, the total FIR luminosity from all galaxies per protocluster is inferred as $L_{\text{FIR}}=1.3^{+1.6}_{-1.0}\times10^{13}L_\odot$. In the case of the SED model without the AGN component, it is estimated as $L_{\text{FIR}}=19.3^{+0.6}_{-4.2}\times10^{13}L_\odot$. As mentioned in \citet{2019ApJ...887..214K}, the best-fit $L_{\text{FIR}}$ has degeneracy between two cases, so the uncertainty is quite large. Considering this point and the effective volume of our protoclusters, our estimation of $\rho_{\text{FIR}}$ is consistent with these estimations.
\par For deriving SFRD, We apply the conversion equation from \citet{1998ARA&A..36..189K} to $\rho_{\text{UV}}$ and $\rho_{\text{FIR}}$, as described below;
\begin{equation}
    {\rm SFRD} = 1.73\times10^{-10}\rho_{\text{FIR}}+1.4\times10^{-28}\rho_{\text{UV}}
\end{equation}
\par As a result, our protocluster galaxies is estimated to have the SFRD corresponding to $\log_{10}{\rm SFRD/(M_\odot\ yr^{-1}\ Mpc^{-3})}= 1.54^{+0.16}_{-0.20}\ (1.68^{+0.16}_{-0.17})$ in the case of the Schechter (DPL) function. This value is roughly $\sim 2.5\ {\rm dex}$ higher than that of field galaxies (e.g., $\log_{10}{\rm SFRD/(M_\odot\ yr^{-1}\ Mpc^{-3})}=-1.00\pm0.06$ in \citet{2015ApJ...803...34B}), suggesting that our protocluster regions have active star formation.
\par Previous studies estimate the SFRD of field LBGs by assuming the $\text{IRX}-\beta$ relation of local starburst galaxies in \citet{Meurer99}. For reference, the SFRD of our protocluster members estimated with this IRX-$\beta$ relation is $\log_{10}{\rm SFRD/(M_\odot\ yr^{-1}\ Mpc^{-3})}=1.61^{+0.33}_{-0.45}\ (1.71^{+0.26}_{-0.31})$ in the case of the Schechter (DPL) function, which is consistent with the original result.
\par Next, we estimate the fraction of the cosmic SFRD from progenitors of massive halos ($M_{\text{halo}} >10^{14} {\rm M_\odot}$). We convert the estimated SFRD, which is per unit volume of protocluster, to that per unit of cosmic volume, and divide it by the field SFRD. The field SFRD is taken from \citet{2015ApJ...803...34B} ($\log_{10}{\rm SFRD/(M_\odot\ yr^{-1}\ Mpc^{-3})}=-1.00\pm0.06$). Using other estimates \citep[e.g.,][]{VanderBurg2010, 2009ApJ...705..936B} changes the result by only $\sim0.1$ dex.
\par In addition, our protocluster sample is not complete for all progenitors of halos of $M_{\text{halo}}>10^{14}{\rm M_\odot}$ at $z\sim0$. Some fraction of dark matter halos with overdensity below $4\sigma$ at $z\sim4$ will also evolve into such halos. We can identify such progenitor halos in the simulation of \citet{Toshikawa18,Toshikawa16}. The fraction of halos that can be observed by our protocluster selection with a galaxy overdensity significance greater than $4\sigma$ at $z\sim4$ is about $6.2\pm1.0\%$, suggesting that our sample has a very high purity but low completeness. The fraction of halos can be translated to the fraction of member galaxies based on the overdensity distribution of progenitor halos, which is equivalent to $9.67\pm0.41\%$. Most of the non-observed member galaxies should be hosted by progenitor halos whose overdensity significance is less than $4\sigma$. With a simple assumption that the UVLF of these galaxies is the same as our PC UVLF, we can derive the intrinsic contribution of progenitor of massive halos to the cosmic SFRD by dividing by this completeness. We mention that the shape of PC UVLF depends on the overdensity, but the main difference of the shape is at $M_{\rm UV}< -22$, which does not significantly affect the SFRD measurement. 
\par Moreover, $76\%$ of our protocluster sample are expected to evolve into $M_{\text{halo}}>10^{14}{\rm M_\odot}$ at $z\sim0$ (T18), so we correct the purity by multiplying this ratio. Finally, we estimate that the $9.4^{+4.7}_{-3.4}\%$ ($13.9^{+6.5}_{-4.9}\%$) of the cosmic SFRD occurs in progenitors of massive halos in the case when we use the best-fit of the Schechter function (the DPL function). 
\par We compare this measurement with the prediction from the SAM in \citet{2017ApJ...844L..23C}. They focus on galaxies with $\log{(M_*/{\rm M_\odot})}>8.5$ in progenitors of cluster of $M_{200}>10^{14}{\rm M_\odot}$ at $z\sim0$, and estimate that the contribution of protocluster galaxies is about $24\ (19)\%$ at $z\sim4$ when they use \citet{2015MNRAS.451.2663H} \citep{2013MNRAS.428.1351G} SAM. 
\par
\begin{figure}
    \centering
    \includegraphics[width=8.5cm]{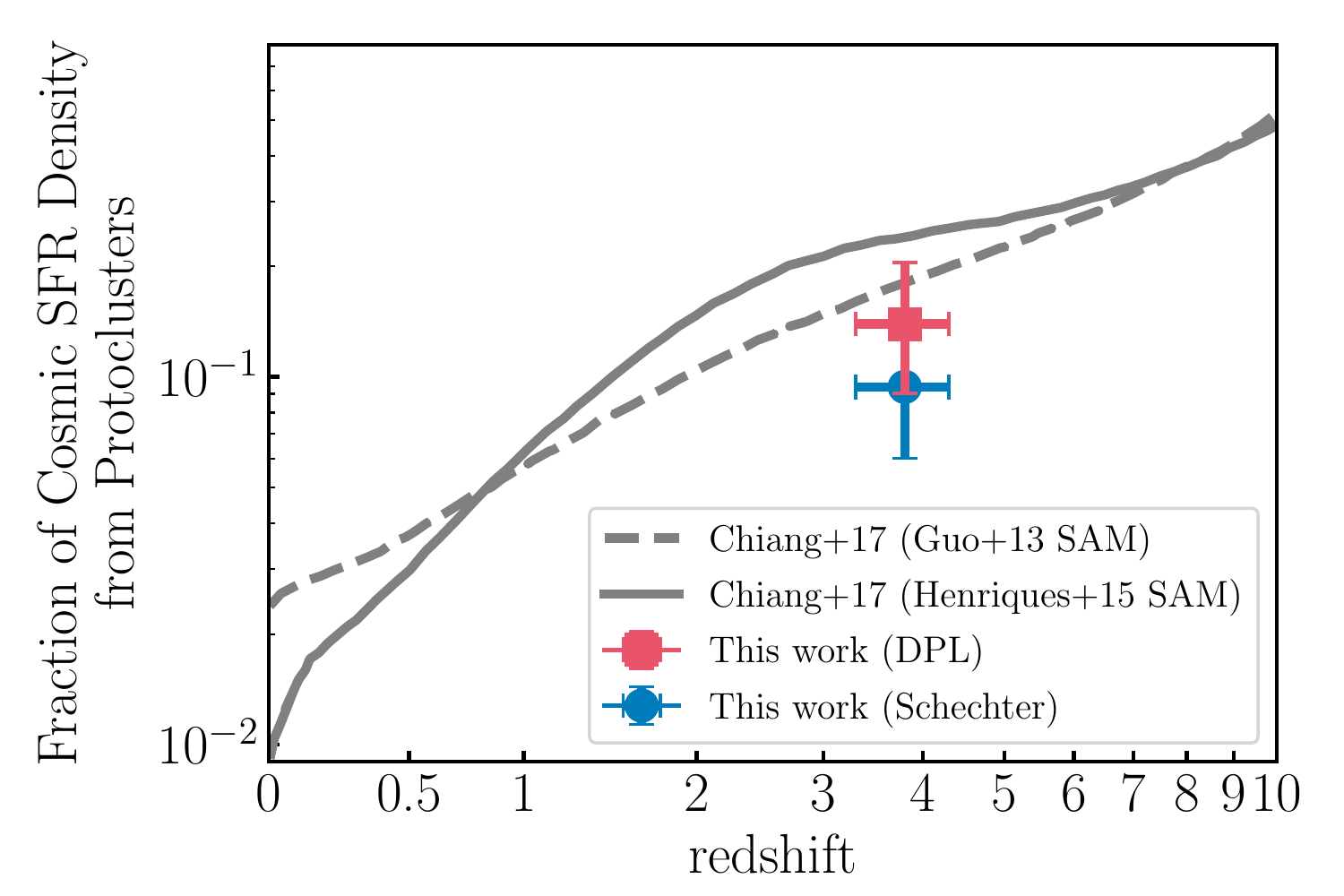}
    \caption{The fraction of the cosmic SFRD in protoclusters. Blue circle and red square represent our estimated value for HSC-SSP protoclusters at $z\sim4$ assuming that PC UVLF follows Schechter function and DPL function, respectively. Gray solid and dashed lines are its predicted evolution in \citet{2017ApJ...844L..23C} with the use of the semi-analytical model of \citet{2015MNRAS.451.2663H} and \citet{2013MNRAS.428.1351G}, respectively.}
    \label{fig:SFRD}
\end{figure}

 The comparison between the observed and predicted fraction of protocluster galaxies to the cosmic SFRD is shown in Figure \ref{fig:SFRD}. Our result is close to the theoretical prediction but slightly smaller. There are two possible explanations. First, we only focus on UV-bright galaxies and miss some other galaxy populations, such as SMGs, which are not selected by LBG selection. Though it is not yet clearly understood how much we miss such galaxies by $g$-dropout selection, \citet{2019Natur.572..211W} argue that the optically-dark but submillimeter-bright galaxies have a significant contribution to the cosmic SFRD. \citet{2018Natur.553...51M} report two SMGs are located in a small separation, implying that they are located in a massive halo. Also, some studies report highly overdense regions of SMGs \citep[e.g.,][]{2018Natur.556..469M}. Although the FIR luminosity galaxies from \citet{2019ApJ...887..214K} has a large degeneracy dependent on the SED model they use, the SFRD combined with the UV luminosity density estimated in this work and the stacked FIR luminosity from \citet{2019ApJ...887..214K} are consistent with the theoretical prediction within the uncertainty. This FIR luminosity, estimated from the stacking, includes the contribution of SMGs, so this does not reject that SMG may be one of the reasons. Second, we may miss some members located on the outskirts of more massive protoclusters. This is because we define protocluster members according to the predicted size of the progenitor of ``Fornax-type" clusters, which can be small for progenitors of more massvie clusters, like ``Coma-like" clusters. 
\par We note that even if we estimate SFRD with the use of the best-fit PC UVLF in the magnitude range of $M_{UV}<-19$, corresponding to $\log{(M_*/{\rm M_\odot})}>8.5$ according to \citet{Song16}, the result does not change significantly. 
\section{Discussion}\label{sec:5}
\subsection{A Possible Confusion Limit}\label{sec:5-1}
\par We have evaluated the sample incompleteness in the same manner as most of the other studies of field LBGs (Section \ref{sec:CF}), and find that it is consistent with previous studies by comparing it with the field UVLF. However, another possible incompleteness could be caused by object confusion in crowded regions, such as in protoclusters. In some overdense regions, some fraction of galaxies will be mixed with nearby objects, which could lower the completeness. Our finding of a flatter UVLF in protoclusters than in the field UVLF could be due to this confusion effect, which might more significantly affect fainter galaxies. The luminosity function shape could change, as seen in this study. 
\par We check this effect by inserting mock galaxies into an overdense region to compare the completeness function in overdense regions with that in the blank field, as estimated in Section \ref{sec:CF}. We summarize the detailed procedure in Appendix \ref{sec:App4} and find that there is no additional incompleteness due to the object confusion in regions with an overdensity significance up to $\sim8\sigma$.
\par Though we see a deficit in the faint-end of the PC UVLF compared to the field UVLF, the cause of this will be further investigated in future studies after we construct a protocluster sample in the Deep layer of HSC-SSP, which has deeper image than the Wide layer. However, we now see that the completeness function of $g$-dropout galaxies estimated in this study is consistent with that of previous studies and that the blending due to focusing on overdense regions like HSC-SSP protoclusters does not lower the completeness. These results imply that the deficit is at least not due to incompleteness.    
\subsection{Spatial Extension of Protoclusters}\label{sec:5-2}
\par We have selected protocluster members from galaxies located within $1.8\arcmin$ from each overdensity peak. Since protoclusters with more significant overdensity tend to be more extended, we may miss some protocluster members on the outskirts of protoclusters, and this could lead to the bright-end excess. To examine this possibility, we redefine protocluster members as galaxies which are located within $4.'2$ from the overdensity peak, which corresponds to the size of progenitors of only the most massive halos ($M_{\text{halo}}>10^{15}{\rm M_\odot}$) like the Coma cluster at $z\sim4$. We find that the shape of PC UVLF does not change from the case of $1.'8$, suggesting that the trend is not caused by the differences in the typical spatial dimensions of protoclusters of different masses. We also check the case of a smaller protocluster radius ($\sim1'$) and find that the trend does not change. 
\subsection{A Possible Excess from the AGN Contribution}\label{sec:5-3}
\par Recent studies have argued that the bright-end ($M_{\text{UV}}\leq-23.0$) of the UVLF at $z\sim4$ is mainly dominated by AGNs \citep[e.g.,][]{2019arXiv191201626A,Ono2018}. Here, we discuss how well the contribution due to the AGNs can explain the bright-end excess that we found in the PC UVLF for $M_{\text{UV}}\leq-20.8$.
\par First of all, we compare our PC UVLF to the field quasar UVLF. \citet{2018PASJ...70S..34A} construct the quasar UVLF at $z\sim4$. The number density of quasars based on the best-fit DPL function for the magnitude range of $-25.8<M_{\text{UV}}<-20.8$, which is the range where our PC UVLF has an excess, is about $(0.9-10)\times10^{-7}\ {\rm \ Mpc^{-3}\ mag^{-1}}$. This value is $(1-240)\times10^{3}$ times lower than the excess at the bright-end that we found in the study. In addition, we have found that UV-luminous quasars scarcely exist in the protoclusters at $z\sim4$ \citep[][in prep.]{2018PASJ...70S..32U}, suggesting that the number density of luminous quasars in protoclusters should not be larger than that in the field.
\par The difference between PC UVLF and the field UVLF in the magnitude range of $ M_{\text{UV}}\leq-20.8$ corresponds to 16 objects per protocluster. The expected total number of members in a protocluster is about 50, indicating that the bright-end excess corresponds to about $32\%$ of the total protocluster members. If we assume that all of the excess at the bright-end is due to the AGN, such a high AGN fraction in protoclusters is inconsistent with previous studies. For example, \citet{Toshikawa16} make follow-up spectroscopy for protocluster member candidates, and they do not find any AGN in 11 members in a protocluster at $z\sim3$, suggesting that the AGN fraction is less than $9\%$. Assuming that the same upper limit for the AGN fraction, the expected number of AGNs in a protocluster is less than five out of 50 members. Other studies show similar AGN fractions for protoclusters from X-ray counterparts. \citet{2009ApJ...691..687L} estimate AGN fraction ($9.5^{+12.7}_{-6.1}$ percent) for LBGs in the SSA22 protocluster at $z=3.09$. \citet{2019ApJ...874...54M} estimate AGN fraction as $2.0^{+2.6}_{-1.3}$ percent for HAEs in the USS1558-003 protocluster at $z=2.53$. \citet{Krishnan17} investigate AGNs in a protocluster called Cl $0218.3-0510$ at $z=1.62$ and estimate that AGN fraction of massive ($\log(M_*/{\rm M_\odot})>10$) protocluster galaxies is $17^{+6}_{-5}$ percent. Though they argue that this value is high compared to that of the blank field at the same redshift, it is not enough to explain the bright-end excess of our PC UVLF. It should be mentioned that the AGN fraction estimated from the X-ray detection can be sensitive to its depth, but these comparison implies that protoclusters at $z\sim4$ are less likely to host such amount of UV-bright AGNs.
\par We note that residuals at $M_{\text{UV}}<-23.0$ of PC UVLF from the best-fit of the Schechter (DPL) correspond to the 1.5/0.5 objects per protocluster. This seems to be reasonable for the AGN fraction in a protocluster; therefore, a part of the bright-end excess can be contributed by the AGN.
\par Therefore, we conclude that AGNs are unlikely to explain all of bright-end excess in the PC UVLF. It should be noted that we here discuss the UV-bright AGNs, and we do not include obscured AGNs. As mentioned in Section \ref{sec:4-1}, \citet{2019ApJ...887..214K} stack IR images of various surveys and estimate the total FIR luminosity of the same protocluster sample with this study. Their results imply that HSC-SSP protoclusters can include a population of UV-dim AGNs.
\subsection{Galaxy Formation in Overdense Regions} \label{sec:5-4}
 \par Some studies suggest that the star formation is enhanced in overdense regions at high-redshift compared to in the blank field, as we mentioned in Section \ref{sec:LF_PC}. For example, HAEs in protoclusters at $z\sim2-2.5$ shows an enhancement of high SFR galaxies \citep{2018MNRAS.473.1977S, 2013MNRAS.428.1551K}. In addition, \citet{2019ApJ...879....9S} report a tentative evidence of higher SFR for Ly$\alpha$ emitting galaxies in protoclusters at $z=3.13$. On the other hand, local galaxy clusters show the opposite trend. For example, cluster galaxies at $0.18<z<0.55$ have SFRs about from $0.00\pm0.11\ h^{-2} {\rm M_\odot}\ {\rm\text{yr}^{-1}}$ to $0.17\pm0.02 \ h^{-2}\ {\rm M_\odot}\ \text{yr}^{-1}$, which are always lower than those of field galaxies \citep{1998ApJ...504L..75B}. Similarly, the low star formation activity in a cluster is also reported at $z=1.6$ \citep{2009A&A...504..331K}. Combining our results with those from the literature, the enhancement of SFR in overdense environments has already started at $z\sim4$, and the star formation activity drops at some time between $z\sim0$ and $z\sim2$, which is earlier than for field galaxies. This is supported by the fact that massive quiescent galaxies have rapidly emerged in overdense regions in the era from $z\sim2.5$ to $z\sim1.5$ \citep[e.g.,][]{2016ApJ...828...56W,2015MNRAS.452.2318C,2014ApJ...788...51N}.
 \par Focusing on the stellar mass, there are several reports that there are more massive galaxies in protoclusters at $z\sim2-3$ \citep{2018MNRAS.473.1977S, 2013MNRAS.428.1551K, 2014MNRAS.440.3262C, 2011MNRAS.415.2993H}, similar to our results at $z\sim4$. At lower redshift ($z<1.5$), the situation is controversial. Many studies report that the shape of the SMFs of star-forming and quiescent galaxies in clusters are similar \citep[e.g.,][]{2017ApJ...851..139L,vanderBurg:2013gl, 2013MNRAS.432.3141C}, while for those of all cluster galaxies, it is argued that there are significant differences not only in the normalization but also in shape at $z\sim1$ in \citet{vanderBurg:2013gl}, at $z\sim0.5-0.7$ in \citet{2018A&A...618A.140V} and at $z\sim0$ in \citet{2001ApJ...557..117B}. In addition, \citet{2010ApJ...718...86K} report a difference between the SMFs of galaxies in a group environment and those in the blank field. On the other hand, \citet{2013MNRAS.432.3141C} suggest that the shape of the SMF is independent of the environment for $z\sim0$, likewise \citet{2016A&A...592A.161N} support for $z\sim1.5$.
 \par It should be noted that some studies report almost no difference from field galaxies in terms of the SFR and stellar mass of protocluster galaxies at $z = 2.9$ \citep{2014A&A...570A..16C}, and at $z=4.57$ \citep{2018A&A...615A..77L}. These studies are based on only spectroscopically confirmed members, which are free from contamination, however the sample of members is small ($\sim 10$ objects), which may not reveal the differences that we find on this study based on the statistical sample. 
 \par These comparisons suggest that galaxies in overdense regions are more massive and have more active star formation compared to galaxies in the blank field at $z>1.5$. Whereas at lower redshift, these trends change; galaxies in overdense regions have lower SFR, and the SMF can be identical to that of the field at least when focusing on the same galaxy population. In addition, star-forming galaxies in protoclusters tend to locate at the main sequence at $z\sim4$ \citep{Long:2020vn,2019ApJ...871...83S}, and $z\sim2-2.5$ \citep{2018MNRAS.473.1977S, 2013MNRAS.428.1551K}. This means that the majority of protocluster members are normal galaxies, and the starburst activity is not significant. Therefore, these results may imply the earlier star formation in protoclusters.
 \par This early formation scenario is consistent with theoretical predictions. \citet{2017ApJ...844L..23C} suggest three phases for the evolution of (proto)clusters. Galaxies in protoclusters already begin star formation in an ``inside-out" manner from $z\geq10$ to $z\sim5$. Then, they continue the star formation from $z\sim5$ to $z\sim1.5$. At $z\leq1.5$, star formation in galaxies is finished, and infalling galaxies into (proto)clusters dominate the main stellar mass growth in protoclusters. Such infalling galaxies are one of the possible reasons that the differences of SMFs of galaxies in local clusters disappear \citep{2013A&A...550A..58V}. Steeper SMFs for cluster galaxies at lower-$z$ seen in Section \ref{sec:4-2} can also be explained by this infalling galaxies effect. In addition, they also imply that $\sim 20\%$ of the cosmic SFRD is contributed by protocluster galaxies, which is roughly consistent with our estimation, as discussed in Section \ref{sec:4-1}.
 \par The shape difference of the PC UVLF and the PC SMF seen in this study can also be related to frequent mergers or an increase in gas supply towards the center of the connection of several connected filaments in an overdense region, as suggested in \citet{2018MNRAS.473.1977S}. Indeed, \citet{2017MNRAS.472.3512T} show that ``top-heavy" SMFs may originate from the enhancement of mergers in overdense regions. They first construct SMFs for star-forming galaxies and quiescent galaxies at $z\sim1$ subdivided by their local environment. They find that shapes of SMFs in more overdense regions tend to be more top-heavy. They try to explain this trend by a simple semi-empirical model. This model first generates $\sim10^6$ galaxies at $z=5$. For each redshift slice, some fraction of galaxy pairs are selected for the merger, and some fraction of galaxies are selected for quenching. The only free parameter is the merged galaxy fraction. The model shows that the observed SMF in overdense regions can be explained by high-merger rate ($80-90\%$). In addition, the increase of gas supply can keep galaxies, which are too massive to be star-forming galaxies in the blank field, to have star-formation. This effect also makes the SMF of protocluster galaxies, which consist only of star-forming ones, to be top-heavy.
\par We find in Section \ref{sec:4-3} that all protoclusters follow the same trend that galaxies in more massive overdense regions tend to have a flatter UVLF, though the diversity exists even if we focus on protoclusters with the same overdensity. The trend implies that more massive regions have generally experienced the earlier structure formation, but their evolutionary stage has a significant variation even at the same epoch. This indicates that a large sample at each redshift is critically essential for tracing the general evolutionary sequence of protoclusters within this diversity.  
\section{Conclusion}\label{sec:6}
\par In this paper, we report the rest-UV luminosity function of $g$-dropout galaxies in 177 protocluster candidates (PC UVLF) at $z\sim4$ detected in the HSC-SSP data. The PC UVLF is estimated in the magnitude range of $-25.8\leq M_{\rm UV}\leq-20.3$ after subtracting for the contamination from field galaxies.
\begin{enumerate}
    \item Compared to the UVLF of galaxies in the blank field, the PC UVLF has a significant excess towards the bright-end in addition to a higher normalization. The best-fit parameters of both the Schechter functions and DPL functions for the PC UVLF also reveal the shape differences from that of the field. The excess towards the bright-end implies that the SFR of galaxies in overdense regions must have accelerated at $z\geq4$.  
    \item  Assuming that all protocluster galaxies follow the ``main sequence" of star-forming galaxies, we convert the PC UVLF to the SMF. Protocluster galaxies are inferred to have 2.8 times more massive characteristic stellar mass than their field counterparts at the same epoch. We show that protocluster galaxies have to continue their stellar mass growth to match SMFs of (proto)cluster galaxies at lower redshift.
    \item More massive protoclusters tend to have a higher bright-end amplitude in the UVLF, although the variation is seen even if we only focus on protoclusters with the same overdensity. The bright-end excess is ubiquitously seen in most of protoclusters at $z\sim4$.
    \item Protoclusters have the enhanced SFRD as $\log_{10}{\rm SFRD/(M_\odot\ yr^{-1}\ Mpc^{-3})}= 1.54^{+0.16}_{-0.20}\ (1.68^{+0.16}_{-0.17})$ using the best-fit of Schechter (DPL) function. This corresponds to the $6-20 \%$ of the cosmic SFRD, being close to the theoretical prediction of \citet{2017ApJ...844L..23C}, but somewhat smaller. This difference from the prediction might be due to the ignorance of SMG in this study and the missed protocluster members located at the edges of protoclusters.
\end{enumerate}
\par Highly star forming and more massive galaxies in protoclusters are reported in protoclusters at lower redshift. We interpret this trend as a signature of the fact that protoclusters are regions in the cosmic web where galaxies and structures form earlier.
\par In this paper, we only focus on protoclusters at $z\sim4$. Currently, we are in the process of selecting protocluster candidates at $z\sim2-6$ from HSC-SSP data in the systematic way same as in T18. This will enable us to determine the UVLF and SFRD of protocluster galaxies at different redshifts and hence trace their redshift evolution.
\\
\par We acknowledge Dr. Masao Hayashi for providing the instruction for the detection of mock galaxies by using {\tt hscpipe}. Also, we appreciate the anonymous referee for helpful comments and suggestions that improved the manuscript.
\par This work was partially supported by Overseas Travel Fund for Students (2019) of the Department of Astronomical Science, the Graduate University for Advanced Studies (SOKENDAI). 
\par The Hyper Suprime-Cam (HSC) collaboration includes the astronomical communities of Japan and Taiwan, and Princeton University. The HSC instrumentation and software were developed by the National Astronomical Observatory of Japan (NAOJ), the Kavli Institute for the Physics and Mathematics of the Universe (Kavli IPMU), the University of Tokyo, the High Energy Accelerator Research Organization (KEK), the Academia Sinica Institute for Astronomy and Astrophysics in Taiwan (ASIAA), and Princeton University. Funding was contributed by the FIRST program from the Japanese Cabinet Office, the Ministry of Education, Culture, Sports, Science and Technology (MEXT), the Japan Society for the Promotion of Science (JSPS), Japan Science and Technology Agency (JST), the Toray Science Foundation, NAOJ, Kavli IPMU, KEK, ASIAA, and Princeton University.
\par This paper makes use of software developed for the Large Synoptic Survey Telescope. We thank the LSST Project for making their code available as free software at {\tt http://dm.lsst.org}.
\par This paper is based on data collected at the Subaru Telescope and retrieved from the HSC data archive system, which is operated by Subaru Telescope and Astronomy Data Center (ADC) at NAOJ. Data analysis was in part carried out with the cooperation of Center for Computational Astrophysics (CfCA), NAOJ.
\par The Pan-STARRS1 Surveys (PS1) and the PS1 public science archive have been made possible through contributions by the Institute for Astronomy, the University of Hawaii, the Pan-STARRS Project Office, the Max Planck Society and its participating institutes, the Max Planck Institute for Astronomy, Heidelberg, and the Max Planck Institute for Extraterrestrial Physics, Garching, The Johns Hopkins University, Durham University, the University of Edinburgh, the Queen???s University Belfast, the Harvard-Smithsonian Center for Astrophysics, the Las Cumbres Observatory Global Telescope Network Incorporated, the National Central University of Taiwan, the Space Telescope Science Institute, the National Aeronautics and Space Administration under grant No. NNX08AR22G issued through the Planetary Science Division of the NASA Science Mission Directorate, the National Science Foundation grant No. AST-1238877, the University of Maryland, Eotvos Lorand University (ELTE), the Los Alamos National Laboratory, and the Gordon and Betty Moore Foundation.
\facilities{Subaru (HSC)}
\appendix
\section{The impact of the photometric uncertainty on the shape of the PC UVLF}\label{sec:App1}
\par Here, we examine the effect of the photometric uncertainty on the shape of the UVLF.
\par
\begin{figure}
    \centering
    \includegraphics[width=15cm]{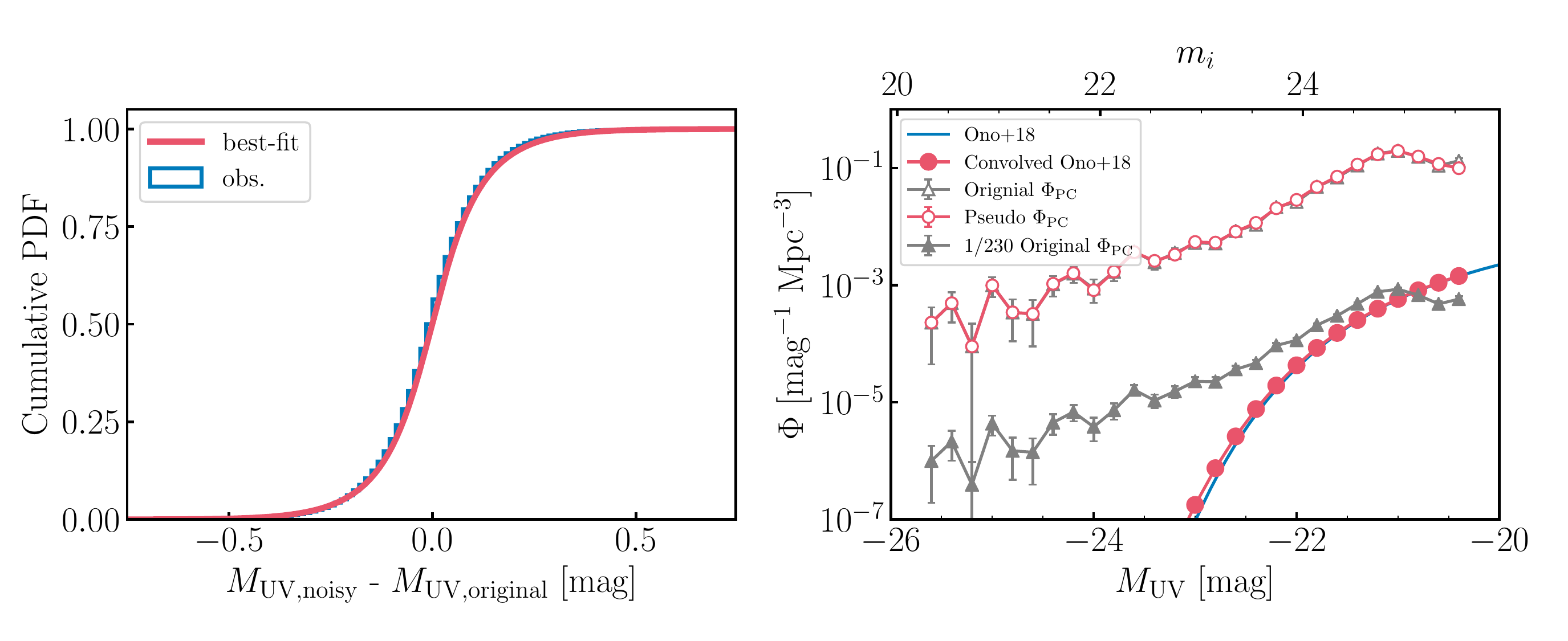}
    \caption{Left panel: The distribution of the difference between the original magnitude and the noise added one, which is shown as blue histogram. The best-fit of the product of the Gaussian distribution and the Lorenzian distribution is shown in the red line. Right panel: The convolved (red filled circles) and original (blue line) field UVLF \citep{Ono2018} and the shifted (gray triangles), pseudo (red open circles), and original (gray open triangles) PC UVLF in this work.}
    \label{fig:App1}
\end{figure}

 We first check the uncertainty of the PC UVLF due to the photometric uncertainty. For each galaxy, we generate mock $M_{\rm UV}$ by adding Gaussian noise whose $1\sigma$ corresponds to the observed photometric error to the observed magnitude. The PC UVLF is recalculated from this $M_{\rm UV}$ distribution with the 1000 times iteration. The right panel of Figure \ref{fig:App1} shows the recalculated PC UVLF (called pseudo PC UVLF) compared with the original PC UVLF. These two UVLFs are consistent, so this implies that the uncertainty of the PC UVLF due to the photometric uncertainty is negligible.
\par We then assess the Eddington Bias. We first derive the difference between the original magnitude and that with artificially noise, estimated in the previous paragraph. Followed by the method in the previous works \citep[e.g.,][]{2013A&A...556A..55I} which estimate the effect of the Eddington Bias to the stellar mass function, the product of the Gaussian distribution $G(x)=\frac{1}{\sigma \sqrt{2 \pi}} \exp{(-\frac{1}{2} \frac{x^{2}}{\sigma^2})}$ and the Lorentzian distribution $L(x)=\frac{\tau}{2 \pi} \frac{1}{\left(\frac{\tau}{2}\right)^{2}+x^{2}}$ is fitted to the magnitude difference distribution, which is shown in the left panel of Figure \ref{fig:App1}, and we obtain the best-fit parameters ($\sigma$ and $\tau$). Convolving the observed field UVLF with the best-fit functions provides us how significant the Eddington bias is in our photometry quality. Here, we employ the best-fit Schechter function of the field UVLF obtained in \citet{Ono2018}.
\par The right panel of Figure \ref{fig:App1} shows the convolved field UVLF. Compared with the original field UVLF, it has indeed slightly higher amplitude than the original one, but it has still steep shape than our estimated PC UVLF. This implies that the our photometric quality does not make the bright-end excess seen in the PC UVLF from the field UVLF.
\section{The PC UVLF in the case of $F(M_{\rm UV})=1$}\label{sec:App2}
\par
\begin{figure}
    \centering
    \includegraphics[width=8cm]{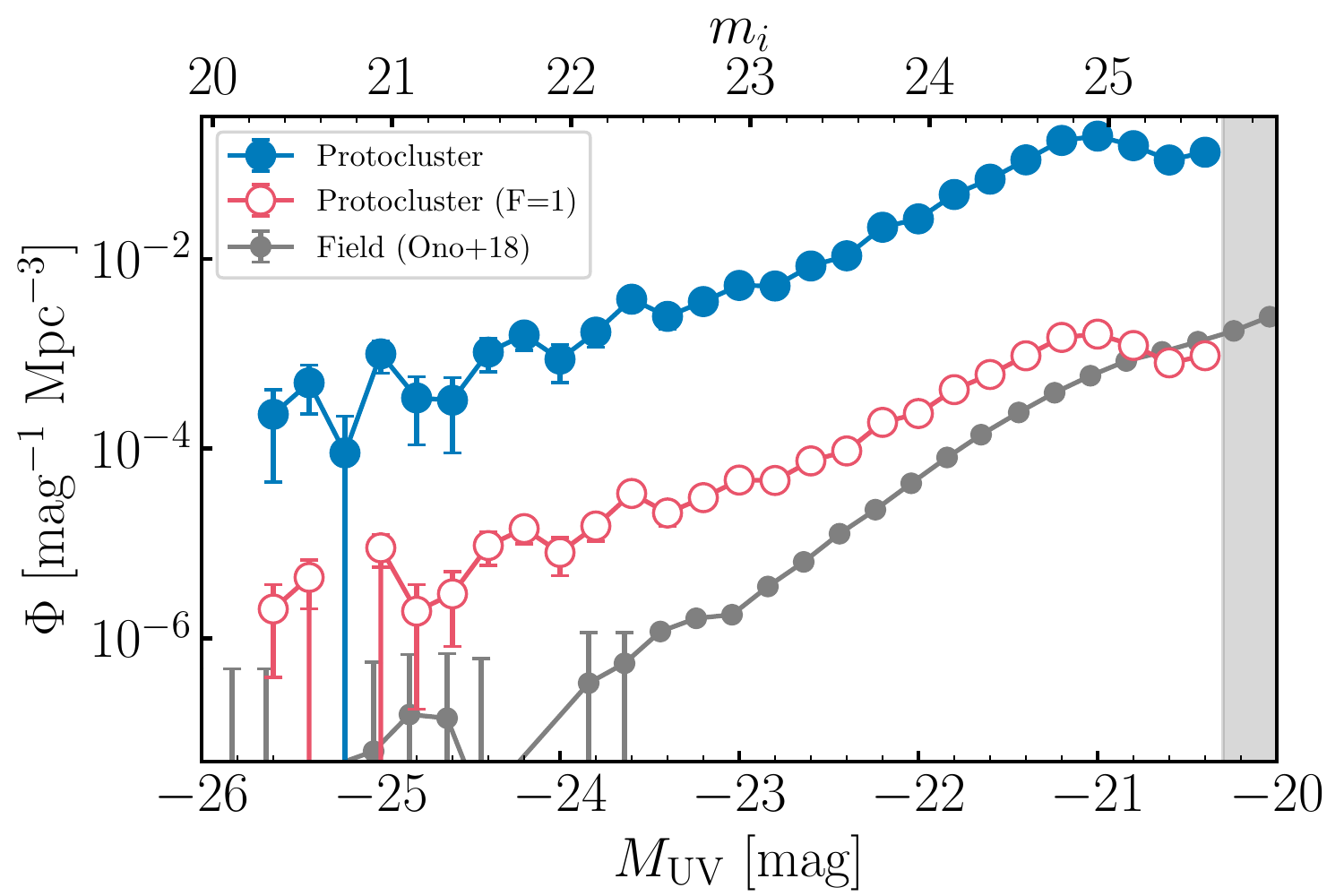}
    \caption{The PC UVLF in the case of $F(M_{\rm UV})=1$ (open red circles) and the original PC UVLF (filled blue circles). For comparison, the field UVLF \citep{Ono2018} is plotted in a gray line.}
    \label{fig:LF_PC_F1}
\end{figure}

 We compare the PC UVLF when we set the volume ratio factor $F(M_{\rm UV})=1$ in Equation \ref{eq:PCLF} with the PC UVLF and the field UVLF \citep{Ono2018}. This PC UVLF still has the bright-end excess compared to the field UVLF, as seen in Figure \ref{fig:LF_PC_F1}.
\section{The robustness of UV slope-magnitude relation}\label{sec:App3}
\par
\begin{figure}
    \centering
    \includegraphics[width=10cm]{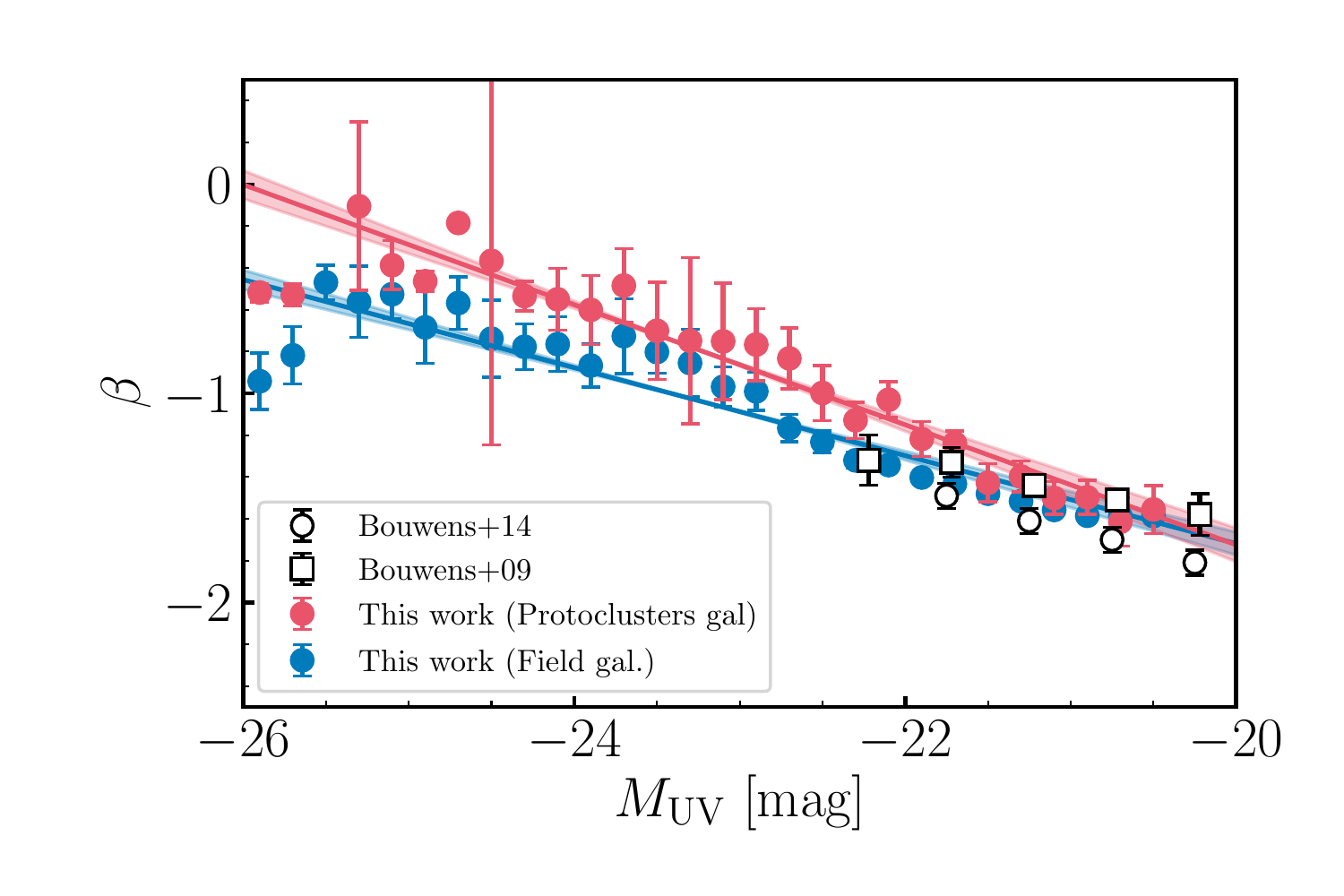}
    \caption{The $\beta-M_{\text{UV}}$ relations estimated in this study. The red circles represent the median value of UV slope of protocluster galaxies and the red line represents its best-fit. The blue circles and the blue line represent those of field galaxies. The shaded regions of each best-fit lines represent their $1\sigma$ uncertainty. Relations from the  literature \citep{2014ApJ...793..115B,2009ApJ...705..936B} is also shown, our estimation for field galaxies is consistent with it, suggesting the robustness of our UV slope estimation.}
    \label{fig:app3}
\end{figure}

 To assess the robustness of our measurement of UV-slope $\beta$ and the sample selection, we measure the $\beta$ of field galaxies in the same manner as described in Section \ref{sec:4-3} to compare it with the relation in the literature. The $\beta-M_{\text{UV}}$ relation our field galaxies as well as in the literature \citep{2014ApJ...793..115B,2009ApJ...705..936B} are shown in Figure \ref{fig:app3}. Our $\beta-M_{\text{UV}}$ relation for field galaxies is consistent with the literature at $-22.3<M_{\rm UV}<-20.3$, suggesting that our measurement and the sample selection is robust.
\section{completeness affected by a possible confusion limit at overdense regions }\label{sec:App4}
\par In this paper, we focus on overdense regions of galaxies. In such regions, the image blending of galaxies might frequently occur due to the high local number density and this could lead an inaccurate photometry of galaxies. The blending also could decrease the sample completeness. This effect is closely related to the number density of galaxies that we focus on. Here, we examine how significantly the blending effect affect the photometry and the completeness by inserting mock galaxies on the image to make artificial overdense regions.
\par Firstly, we make a cut out image with $4\arcmin\times4\arcmin$ of a overdense region whose overdensity peak is about $3\sigma$. In \citet{Toshikawa18}, the average and the standard deviation of the number of bright ($m_i<25$ mag) galaxies within $1'.8$ are $6.4$ and $3.2$, respectively. According to the field luminosity function of $g$-dropout galaxies \citep[e.g.,][]{Ono2018}, this implies that $1\sigma$ of the number density of galaxies with $25<m_i<26$ is about $1.8\ {\rm mag^{-1}arcmin^{-2}}$. We make five artificial overdense region image by inserting mock galaxies to the cutout image so that their number densities are equivalent to $1.8,\ 3.6,\ 5.4,\ 7.2$, and $9.0 \ {\rm arcmin^{-2}mag^{-1}}$, corresponding to the overdensity significance of $4\sigma,\ 5\sigma,\ 6\sigma,\ 7\sigma,\ {\rm and}\ 8\sigma$, respectively, which is the same overdensity range of our protocluster sample. The morphological and physical properties of mock galaxies are the same as that we did in estimating the completeness in Section \ref{sec:CF}. We fix the redshift as $z=3.8$ since we only aim to see the difference induced by the number density of galaxies in the field. The detection and the measurement process are also the same as that described in Section \ref{sec:CF}.
\par
\begin{figure}
    \centering
    \includegraphics[width=12cm]{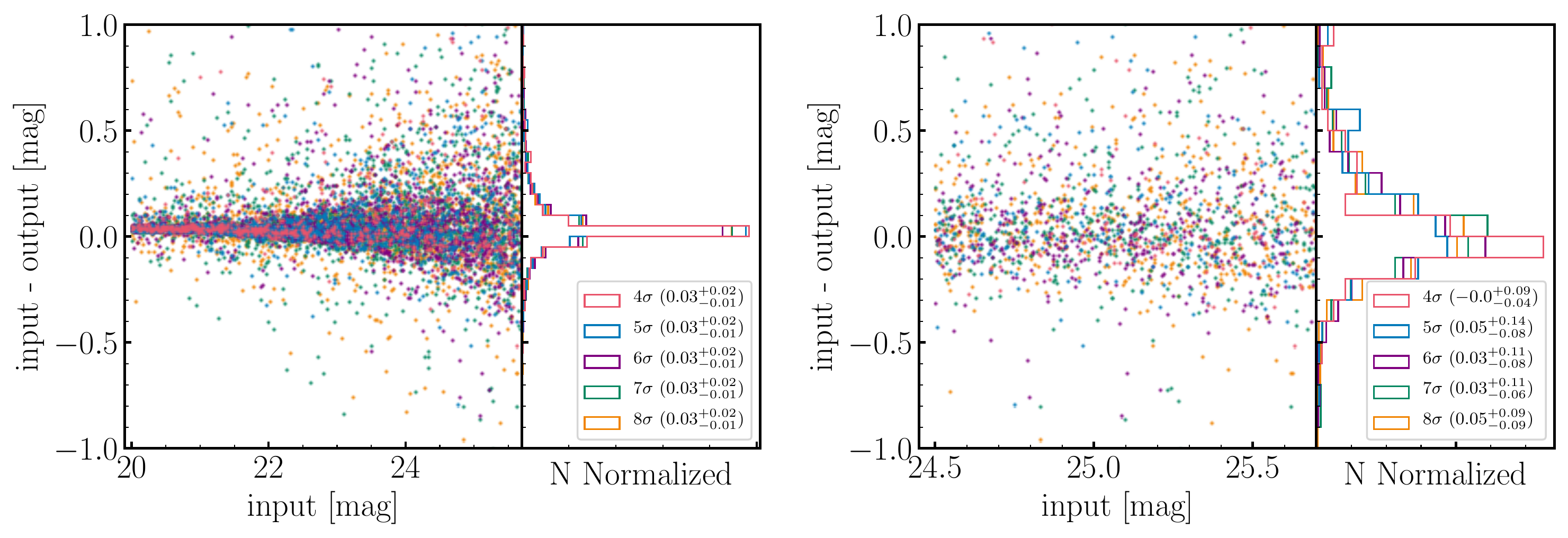}
    \caption{The comparison of the input magnitude and the output magnitude of mock galaxies. Red, blue, purple, green, and yellow markers show cases in $4\sigma$, $5\sigma$, $6\sigma$, $7\sigma$, $8\sigma$ regions. The left panel plots all detected mock galaxies, and the right panel plots only faint galaxies with $m_i>24.5$. In each panel, the median value and 16/84th percentile uncertainty is shown, and all of them is consistent, suggesting that the blending due to the overdensity does not affect the photometry.}
    \label{fig:phot}
\end{figure}

 We compare the output magnitude of the detected objects from {\tt hscpipe} to the input magnitude of mock galaxies in Figure \ref{fig:phot}. The magnitude difference between the input and output magnitude are consistent at any overdensities. The peak difference between the input and the output magnitude is lower than the photometric error, suggesting that the magnitudes are accurately recovered. This result implies that the photometry is not affected by the blending due to the overdensity. Even we only focus on faint ($m_i>24.5$) objects, which can be more blended by other bright objects, they also follow the same trend (the right panel of Figure \ref{fig:phot}).
\par
\begin{figure}
    \centering
    \includegraphics[width=8.5cm]{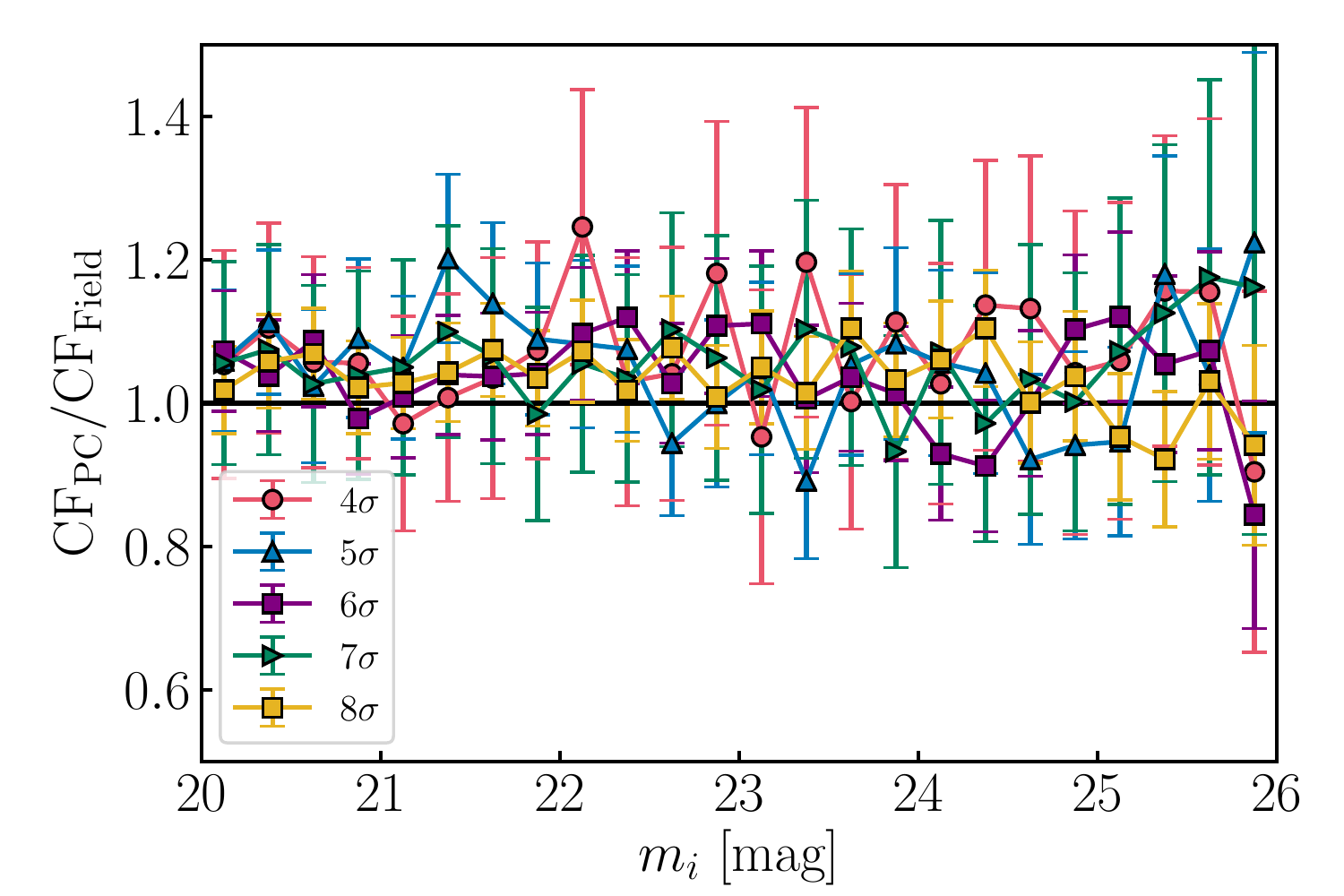}
    \caption{The ratio of the completeness function in overdense regions to that in the blank field. Red, blue, purple, green, and yellow lines correspond to cases in  $4\sigma$, $5\sigma$, $6\sigma$, $7\sigma$, $8\sigma$ regions, respectively. Error bars represent the Poisson error of the number of detected mock galaxies for each bins.}
    \label{fig:OS_CF}
\end{figure}

 As same in Section \ref{sec:CF}, we construct a completeness function as a function of magnitude.  Figure \ref{fig:OS_CF} shows the ratio between these completeness function and that at $z=3.8$ estimated in Section \ref{sec:CF}. The ratio do not change around one up to $8\sigma$. This suggests that the overdensity in the range of that of our protoclusters does not affect the completeness function. 

\bibliographystyle{aasjournal}

\begin{thebibliography}{}
\expandafter\ifx\csname natexlab\endcsname\relax\def\natexlab#1{#1}\fi
\providecommand{\url}[1]{\href{#1}{#1}}
\providecommand{\dodoi}[1]{doi:~\href{http://doi.org/#1}{\nolinkurl{#1}}}
\providecommand{\doeprint}[1]{\href{http://ascl.net/#1}{\nolinkurl{http://ascl.net/#1}}}
\providecommand{\doarXiv}[1]{\href{https://arxiv.org/abs/#1}{\nolinkurl{https://arxiv.org/abs/#1}}}

\bibitem[{Adams {et~al.}(2019)Adams, Bowler, Jarvis, H{\"a}ussler, McLure,
  Bunker, Dunlop, \& Verma}]{2019arXiv191201626A}
Adams, N.~J., Bowler, R. A.~A., Jarvis, M.~J., {et~al.} 2019, arXiv.org,
  arXiv:1912.01626

\bibitem[{Aihara {et~al.}(2018{\natexlab{a}})Aihara, Arimoto, Armstrong,
  Arnouts, Bahcall, Bickerton, Bosch, Bundy, Capak, Chan, Chiba, Coupon, Egami,
  Enoki, Finet, Fujimori, Fujimoto, Furusawa, Furusawa, Goto, Goulding, Greco,
  Greene, Gunn, Hamana, Harikane, Hashimoto, Hattori, Hayashi, Hayashi,
  He{\l}miniak, Higuchi, Hikage, Ho, Hsieh, Huang, Huang, Ikeda, Imanishi,
  Inoue, Iwasawa, Iwata, Jaelani, Jian, Kamata, Karoji, Kashikawa, Katayama,
  Kawanomoto, Kayo, Koda, Koike, Kojima, Komiyama, Konno, Koshida, Koyama,
  Kusakabe, Leauthaud, Lee, Lin, Lin, Lupton, Mandelbaum, Matsuoka, Medezinski,
  Mineo, Miyama, Miyatake, Miyazaki, Momose, More, More, Moritani, Moriya,
  Morokuma, Mukae, Murata, Murayama, Nagao, Nakata, Niida, Niikura, Nishizawa,
  Obuchi, Oguri, Oishi, Okabe, Okamoto, Okura, Ono, Onodera, Onoue, Osato,
  Ouchi, Price, Pyo, Sako, Sawicki, Shibuya, Shimasaku, Shimono, Shirasaki,
  Silverman, Simet, Speagle, Spergel, Strauss, Sugahara, Sugiyama, Suto, Suyu,
  Suzuki, Tait, Takada, Takata, Tamura, Tanaka, Tanaka, Tanaka, Tanaka, Terai,
  Terashima, Toba, Tominaga, Toshikawa, Turner, Uchida, Uchiyama, Umetsu,
  Uraguchi, Urata, Usuda, Utsumi, Wang, Wang, Wong, Yabe, Yamada, Yamanoi,
  Yasuda, Yeh, Yonehara, \& Yuma}]{Aihara18a}
Aihara, H., Arimoto, N., Armstrong, R., {et~al.} 2018{\natexlab{a}}, \pasj, 70,
  S4

\bibitem[{Aihara {et~al.}(2018{\natexlab{b}})Aihara, Armstrong, Bickerton,
  Bosch, Coupon, Furusawa, Hayashi, Ikeda, Kamata, Karoji, Kawanomoto, Koike,
  Komiyama, Lang, Lupton, Mineo, Miyatake, Miyazaki, Morokuma, Obuchi, Oishi,
  Okura, Price, Takata, Tanaka, Tanaka, Tanaka, Uchida, Uraguchi, Utsumi, Wang,
  Yamada, Yamanoi, Yasuda, Arimoto, Chiba, Finet, Fujimori, Fujimoto, Furusawa,
  Goto, Goulding, Gunn, Harikane, Hattori, Hayashi, He{\l}miniak, Higuchi,
  Hikage, Ho, Hsieh, Huang, Huang, Imanishi, Iwata, Jaelani, Jian, Kashikawa,
  Katayama, Kojima, Konno, Koshida, Kusakabe, Leauthaud, Lee, Lin, Lin,
  Mandelbaum, Matsuoka, Medezinski, Miyama, Momose, More, More, Mukae, Murata,
  Murayama, Nagao, Nakata, Niida, Niikura, Nishizawa, Oguri, Okabe, Ono,
  Onodera, Onoue, Ouchi, Pyo, Shibuya, Shimasaku, Simet, Speagle, Spergel,
  Strauss, Sugahara, Sugiyama, Suto, Suzuki, Tait, Takada, Terai, Toba, Turner,
  Uchiyama, Umetsu, Urata, Usuda, Yeh, \& Yuma}]{Aihara18b}
Aihara, H., Armstrong, R., Bickerton, S., {et~al.} 2018{\natexlab{b}}, \pasj,
  70, S8

\bibitem[{Akiyama {et~al.}(2018)Akiyama, He, Ikeda, Niida, Nagao, Bosch,
  Coupon, Enoki, Imanishi, Kashikawa, Kawaguchi, Komiyama, Lee, Matsuoka,
  Miyazaki, Nishizawa, Oguri, Ono, Onoue, Ouchi, Schulze, Silverman, Tanaka,
  Tanaka, Terashima, Toba, \& Ueda}]{2018PASJ...70S..34A}
Akiyama, M., He, W., Ikeda, H., {et~al.} 2018, \pasj, 70, S34

\bibitem[{Alam {et~al.}(2015)Alam, Albareti, Allende~Prieto, Anders, Anderson,
  Anderton, Andrews, Armengaud, Aubourg, Bailey, Basu, Bautista, Beaton, Beers,
  Bender, Berlind, Beutler, Bhardwaj, Bird, Bizyaev, Blake, Blanton, Blomqvist,
  Bochanski, Bolton, Bovy, Shelden~Bradley, Brandt, Brauer, Brinkmann, Brown,
  Brownstein, Burden, Burtin, Busca, Cai, Capozzi, Carnero~Rosell, Carr,
  Carrera, Chambers, Chaplin, Chen, Chiappini, Chojnowski, Chuang, Clerc,
  Comparat, Covey, Croft, Cuesta, Cunha, da~Costa, Da~Rio, Davenport, Dawson,
  De~Lee, Delubac, Deshpande, Dhital, Dutra-Ferreira, Dwelly, Ealet, Ebelke,
  Edmondson, Eisenstein, Ellsworth, Elsworth, Epstein, Eracleous, Escoffier,
  Esposito, Evans, Fan, Fern{\'a}ndez-Alvar, Feuillet, Filiz~Ak, Finley,
  Finoguenov, Flaherty, Fleming, Font-Ribera, Foster, Frinchaboy,
  Galbraith-Frew, Garc{\'\i}a, Garc{\'\i}a-Hern{\'a}ndez,
  Garc{\'\i}a~P{\'e}rez, Gaulme, Ge, G{\'e}nova-Santos, Georgakakis, Ghezzi,
  Gillespie, Girardi, Goddard, Gontcho, Gonz{\'a}lez~Hern{\'a}ndez, Grebel,
  Green, Grieb, Grieves, Gunn, Guo, Harding, Hasselquist, Hawley, Hayden,
  Hearty, Hekker, Ho, Hogg, Holley-Bockelmann, Holtzman, Honscheid, Huber,
  Huehnerhoff, Ivans, Jiang, Johnson, Kinemuchi, Kirkby, Kitaura, Klaene,
  Knapp, Kneib, Koenig, Lam, Lan, Lang, Laurent, Le~Goff, Leauthaud, Lee, Lee,
  Licquia, Liu, Long, L{\'o}pez-Corredoira, Lorenzo-Oliveira, Lucatello,
  Lundgren, Lupton, Mack, Mahadevan, Maia, Majewski, Malanushenko,
  Malanushenko, Manchado, Manera, Mao, Maraston, Marchwinski, Margala, Martell,
  Martig, Masters, Mathur, McBride, McGehee, McGreer, McMahon, M{\'e}nard,
  Menzel, Merloni, M{\'e}sz{\'a}ros, Miller, Miralda-Escud{\'e}, Miyatake,
  Montero-Dorta, More, Morganson, Morice-Atkinson, Morrison, Mosser, Muna,
  Myers, Nandra, Newman, Neyrinck, Nguyen, Nichol, Nidever, Noterdaeme, Nuza,
  O'Connell, O'Connell, O'Connell, Ogando, Olmstead, Oravetz, Oravetz, Osumi,
  Owen, Padgett, Padmanabhan, Paegert, Palanque-Delabrouille, Pan, Parejko,
  P{\^a}ris, Park, Pattarakijwanich, Pellejero-Ibanez, Pepper, Percival,
  P{\'e}rez-Fournon, P~rez Ra~fols, Petitjean, Pieri, Pinsonneault, Porto~de
  Mello, Prada, Prakash, Price-Whelan, Protopapas, Raddick, Rahman, \&
  Reid...}]{2015ApJS..219...12A}
Alam, S., Albareti, F.~D., Allende~Prieto, C., {et~al.} 2015, \apjs, 219, 12

\bibitem[{{\'A}lvarez-M{\'a}rquez {et~al.}(2019){\'A}lvarez-M{\'a}rquez,
  Burgarella, Buat, Ilbert, \& P{\'e}rez-Gonz{\'a}lez}]{2019A&A...630A.153A}
{\'A}lvarez-M{\'a}rquez, J., Burgarella, D., Buat, V., Ilbert, O., \&
  P{\'e}rez-Gonz{\'a}lez, P.~G. 2019, \aap, 630, A153

\bibitem[{Axelrod {et~al.}(2010)Axelrod, Kantor, Lupton, \&
  Pierfederici}]{Axelrod10}
Axelrod, T., Kantor, J., Lupton, R.~H., \& Pierfederici, F. 2010, in
  Proceedings of the SPIE, ed. N.~M. Radziwill \& A.~Bridger, Steward
  Observatory, United States (SPIE), 774015

\bibitem[{Balogh {et~al.}(2001)Balogh, Christlein, Zabludoff, \&
  Zaritsky}]{2001ApJ...557..117B}
Balogh, M.~L., Christlein, D., Zabludoff, A.~I., \& Zaritsky, D. 2001, \apj,
  557, 117

\bibitem[{Balogh {et~al.}(1998)Balogh, Schade, Morris, Yee, Carlberg, \&
  Ellingson}]{1998ApJ...504L..75B}
Balogh, M.~L., Schade, D., Morris, S.~L., {et~al.} 1998, \apj, 504, L75

\bibitem[{Bamford {et~al.}(2008)Bamford, Nichol, Baldry, Land, Lintott,
  Schawinski, Slosar, Szalay, Thomas, Torki, Andreescu, Edmondson, Miller,
  Murray, Raddick, \& Vandenberg}]{Bamford09}
Bamford, S.~P., Nichol, R.~C., Baldry, I.~K., {et~al.} 2008, arXiv.org, 1324

\bibitem[{Bertin(2011)}]{2011ASPC..442..435B}
Bertin, E. 2011, in Astronomical Data Analysis Software and Systems XX. ASP
  Conference Proceedings, 435--

\bibitem[{Biviano {et~al.}(2017)Biviano, Moretti, Paccagnella, Poggianti,
  Bettoni, Gullieuszik, Vulcani, Fasano, D'Onofrio, Fritz, \&
  Cava}]{2017A&A...607A..81B}
Biviano, A., Moretti, A., Paccagnella, A., {et~al.} 2017, \aap, 607, A81

\bibitem[{Boquien {et~al.}(2019)Boquien, Burgarella, Roehlly, Buat, Ciesla,
  Corre, Inoue, \& Salas}]{2019A&A...622A.103B}
Boquien, M., Burgarella, D., Roehlly, Y., {et~al.} 2019, \aap, 622, A103

\bibitem[{Bosch {et~al.}(2018)Bosch, Armstrong, Bickerton, Furusawa, Ikeda,
  Koike, Lupton, Mineo, Price, Takata, Tanaka, Yasuda, AlSayyad, Becker,
  Coulton, Coupon, Garmilla, Huang, Krughoff, Lang, Leauthaud, Lim, Lust,
  MacArthur, Mandelbaum, Miyatake, Miyazaki, Murata, More, Okura, Owen,
  Swinbank, Strauss, Yamada, \& Yamanoi}]{Bosch18}
Bosch, J., Armstrong, R., Bickerton, S., {et~al.} 2018, \pasj, 70, S5

\bibitem[{Bouwens {et~al.}(2009)Bouwens, Illingworth, Franx, Chary, Meurer,
  Conselice, Ford, Giavalisco, \& van Dokkum}]{2009ApJ...705..936B}
Bouwens, R.~J., Illingworth, G.~D., Franx, M., {et~al.} 2009, \apj, 705, 936

\bibitem[{Bouwens {et~al.}(2012)Bouwens, Illingworth, Oesch, Franx, Labb{\'e},
  Trenti, van Dokkum, Carollo, Gonzalez, Smit, \& Magee}]{2012ApJ...754...83B}
Bouwens, R.~J., Illingworth, G.~D., Oesch, P.~A., {et~al.} 2012, \apj, 754, 83

\bibitem[{Bouwens {et~al.}(2014)Bouwens, Illingworth, Oesch, Labb{\'e}, van
  Dokkum, Trenti, Franx, Smit, Gonzalez, \& Magee}]{2014ApJ...793..115B}
---. 2014, \apj, 793, 115

\bibitem[{Bouwens {et~al.}(2015)Bouwens, Illingworth, Oesch, Trenti, Labb{\'e},
  Bradley, Carollo, van Dokkum, Gonzalez, Holwerda, Franx, Spitler, Smit, \&
  Magee}]{2015ApJ...803...34B}
---. 2015, \apj, 803, 34

\bibitem[{Bowler {et~al.}(2019)Bowler, Jarvis, Dunlop, McLure, McLeod, Adams,
  Milvang-Jensen, \& McCracken}]{Bowler:2019tb}
Bowler, R. A.~A., Jarvis, M.~J., Dunlop, J.~S., {et~al.} 2019, arXiv.org

\bibitem[{Bowler {et~al.}(2015)Bowler, Dunlop, McLure, McCracken,
  Milvang-Jensen, Furusawa, Taniguchi, Le~Fevre, Fynbo, Jarvis, \&
  H{\"a}ussler}]{2015MNRAS.452.1817B}
Bowler, R. A.~A., Dunlop, J.~S., McLure, R.~J., {et~al.} 2015, \mnras, 452,
  1817

\bibitem[{Bruzual \& Charlot(2003)}]{Bruzual03}
Bruzual, G., \& Charlot, S. 2003, \mnras, 344, 1000

\bibitem[{Cai {et~al.}(2016)Cai, Fan, Peirani, Bian, Frye, McGreer, Prochaska,
  Lau, Tejos, Ho, \& Schneider}]{2016ApJ...833..135C}
Cai, Z., Fan, X., Peirani, S., {et~al.} 2016, \apj, 833, 135

\bibitem[{Calvi {et~al.}(2013)Calvi, Poggianti, Vulcani, \&
  Fasano}]{2013MNRAS.432.3141C}
Calvi, R., Poggianti, B.~M., Vulcani, B., \& Fasano, G. 2013, \mnras, 432, 3141

\bibitem[{Calzetti {et~al.}(2000)Calzetti, Armus, Bohlin, Kinney, Koornneef, \&
  Storchi-Bergmann}]{Calzetti00}
Calzetti, D., Armus, L., Bohlin, R.~C., {et~al.} 2000, \apj, 533, 682

\bibitem[{Chabrier(2003)}]{2003ApJ...586L.133C}
Chabrier, G. 2003, \apj, 586, L133

\bibitem[{Chiang {et~al.}(2013)Chiang, Overzier, \&
  Gebhardt}]{2013ApJ...779..127C}
Chiang, Y.-K., Overzier, R.~A., \& Gebhardt, K. 2013, \apj, 779, 127

\bibitem[{Chiang {et~al.}(2014)Chiang, Overzier, \& Gebhardt}]{Chiang14}
---. 2014, \apjl, 782, L3

\bibitem[{Chiang {et~al.}(2017)Chiang, Overzier, Gebhardt, \&
  Henriques}]{2017ApJ...844L..23C}
Chiang, Y.-K., Overzier, R.~A., Gebhardt, K., \& Henriques, B. 2017, \apjl,
  844, L23

\bibitem[{Cooke {et~al.}(2014)Cooke, Hatch, Muldrew, Rigby, \&
  Kurk}]{2014MNRAS.440.3262C}
Cooke, E.~A., Hatch, N.~A., Muldrew, S.~I., Rigby, E.~E., \& Kurk, J.~D. 2014,
  \mnras, 440, 3262

\bibitem[{Cooke {et~al.}(2015)Cooke, Hatch, Rettura, Wylezalek, Galametz,
  Stern, Brodwin, Muldrew, Almaini, Conselice, Eisenhardt, Hartley, Jarvis,
  Seymour, \& Stanford}]{2015MNRAS.452.2318C}
Cooke, E.~A., Hatch, N.~A., Rettura, A., {et~al.} 2015, \mnras, 452, 2318

\bibitem[{Cucciati {et~al.}(2012)Cucciati, Tresse, Ilbert, Le~Fevre, Garilli,
  Le~Brun, Cassata, Franzetti, Maccagni, Scodeggio, Zucca, Zamorani, Bardelli,
  Bolzonella, Bielby, McCracken, Zanichelli, \& Vergani}]{2012A&A...539A..31C}
Cucciati, O., Tresse, L., Ilbert, O., {et~al.} 2012, \aap, 539, A31

\bibitem[{Cucciati {et~al.}(2014)Cucciati, Zamorani, Lemaux, Bardelli, Cimatti,
  Le~Fevre, Cassata, Garilli, Le~Brun, Maccagni, Pentericci, Tasca, Thomas,
  Vanzella, Zucca, Amorin, Capak, Cassara, Castellano, Cuby, de~la Torre,
  Durkalec, Fontana, Giavalisco, Grazian, Hathi, Ilbert, Moreau, Paltani,
  Ribeiro, Salvato, Schaerer, Scodeggio, Sommariva, Talia, Taniguchi, Tresse,
  Vergani, Wang, Charlot, Contini, Fotopoulou, Lopez-Sanjuan, Mellier, \&
  Scoville}]{2014A&A...570A..16C}
Cucciati, O., Zamorani, G., Lemaux, B.~C., {et~al.} 2014, \aap, 570, A16

\bibitem[{Davidzon {et~al.}(2017)Davidzon, Ilbert, Laigle, Coupon, McCracken,
  Delvecchio, Masters, Capak, Hsieh, Le~Fevre, Tresse, B{\'e}thermin, Chang,
  Faisst, Le~Floc'h, Steinhardt, Toft, Aussel, Dubois, Hasinger, Salvato,
  Sanders, Scoville, \& Silverman}]{2017A&A...605A..70D}
Davidzon, I., Ilbert, O., Laigle, C., {et~al.} 2017, \aap, 605, A70

\bibitem[{Demarco {et~al.}(2010)Demarco, Wilson, Muzzin, Lacy, Surace, Yee,
  Hoekstra, Blindert, \& Gilbank}]{2010ApJ...711.1185D}
Demarco, R., Wilson, G., Muzzin, A., {et~al.} 2010, \apj, 711, 1185

\bibitem[{Dressler(1980)}]{Dressler80}
Dressler, A. 1980, \apj, 236, 351

\bibitem[{Eddington(1913)}]{1913MNRAS..73..359E}
Eddington, A.~S. 1913, \mnras, 73, 359

\bibitem[{Fasano {et~al.}(2006)Fasano, Marmo, Varela, D'Onofrio, Poggianti,
  Moles, Pignatelli, Bettoni, Kj{\ae}rgaard, Rizzi, Couch, \&
  Dressler}]{2006A&A...445..805F}
Fasano, G., Marmo, C., Varela, J., {et~al.} 2006, \aap, 445, 805

\bibitem[{Finkelstein {et~al.}(2015)Finkelstein, Ryan, Papovich, Dickinson,
  Song, Somerville, Ferguson, Salmon, Giavalisco, Koekemoer, Ashby, Behroozi,
  Castellano, Dunlop, Faber, Fazio, Fontana, Grogin, Hathi, Jaacks, Kocevski,
  Livermore, McLure, Merlin, Mobasher, Newman, Rafelski, Tilvi, \&
  Willner}]{2015ApJ...810...71F}
Finkelstein, S.~L., Ryan, R. E.~J., Papovich, C., {et~al.} 2015, \apj, 810, 71

\bibitem[{Garilli {et~al.}(2014)Garilli, Guzzo, Scodeggio, Bolzonella, Abbas,
  Adami, Arnouts, Bel, Bottini, Branchini, Cappi, Coupon, Cucciati, Davidzon,
  De~Lucia, de~la Torre, Franzetti, Fritz, Fumana, Granett, Ilbert, Iovino,
  Krywult, Le~Brun, Le~Fevre, Maccagni, Ma{\l}ek, Marulli, McCracken, Paioro,
  Polletta, Pollo, Schlagenhaufer, Tasca, Tojeiro, Vergani, Zamorani,
  Zanichelli, Burden, Di~Porto, Marchetti, Marinoni, Mellier, Moscardini,
  Nichol, Peacock, Percival, Phleps, \& Wolk}]{2014A&A...562A..23G}
Garilli, B., Guzzo, L., Scodeggio, M., {et~al.} 2014, Astronomy {\&}
  Astrophysics, 562, A23

\bibitem[{Guo {et~al.}(2013)Guo, White, Angulo, Henriques, Lemson,
  Boylan-Kolchin, Thomas, \& Short}]{2013MNRAS.428.1351G}
Guo, Q., White, S., Angulo, R.~E., {et~al.} 2013, \mnras, 428, 1351

\bibitem[{Hatch {et~al.}(2011)Hatch, Kurk, Pentericci, Venemans, Kuiper, Miley,
  \& R{\"o}ttgering}]{2011MNRAS.415.2993H}
Hatch, N.~A., Kurk, J.~D., Pentericci, L., {et~al.} 2011, \mnras, 415, 2993

\bibitem[{Hayashi {et~al.}(2012)Hayashi, Kodama, Tadaki, Koyama, \&
  Tanaka}]{Hayashi:2012ht}
Hayashi, M., Kodama, T., Tadaki, K.-i., Koyama, Y., \& Tanaka, I. 2012, \apj,
  757, 15

\bibitem[{Henriques {et~al.}(2015)Henriques, White, Thomas, Angulo, Guo,
  Lemson, Springel, \& Overzier}]{2015MNRAS.451.2663H}
Henriques, B. M.~B., White, S. D.~M., Thomas, P.~A., {et~al.} 2015, \mnras,
  451, 2663

\bibitem[{Higuchi {et~al.}(2019)Higuchi, Ouchi, Ono, Shibuya, Toshikawa,
  Harikane, Kojima, Chiang, Egami, Kashikawa, Overzier, Konno, Inoue, Hasegawa,
  Fujimoto, Goto, Ishikawa, Ito, Komiyama, \& Tanaka}]{2019ApJ...879...28H}
Higuchi, R., Ouchi, M., Ono, Y., {et~al.} 2019, \apj, 879, 28

\bibitem[{Hogg(1999)}]{1999astro.ph..5116H}
Hogg, D.~W. 1999, arXiv.org, arXiv:astro

\bibitem[{Ilbert {et~al.}(2013)Ilbert, McCracken, Le~Fevre, Capak, Dunlop,
  Karim, Renzini, Caputi, Boissier, Arnouts, Aussel, Comparat, Guo, Hudelot,
  Kartaltepe, Kneib, Krogager, Le~Floc'h, Lilly, Mellier, Milvang-Jensen,
  Moutard, Onodera, Richard, Salvato, Sanders, Scoville, Silverman, Taniguchi,
  Tasca, Thomas, Toft, Tresse, Vergani, Wolk, \& Zirm}]{2013A&A...556A..55I}
Ilbert, O., McCracken, H.~J., Le~Fevre, O., {et~al.} 2013, \aap, 556, A55

\bibitem[{Ito {et~al.}(2019)Ito, Kashikawa, Toshikawa, Overzier, Tanaka, Kubo,
  Shibuya, Ishikawa, Onoue, Uchiyama, Liang, Higuchi, Martin, Lee, Komiyama, \&
  Huang}]{2019ApJ...878...68I}
Ito, K., Kashikawa, N., Toshikawa, J., {et~al.} 2019, \apj, 878, 68

\bibitem[{Ivezi{\'c} {et~al.}(2008)Ivezi{\'c}, Kahn, Tyson, Abel, Acosta,
  Allsman, Alonso, AlSayyad, Anderson, Andrew, Angel, Angeli, Ansari,
  Antilogus, Araujo, Armstrong, Arndt, Astier, Aubourg, Auza, Axelrod, Bard,
  Barr, Barrau, Bartlett, Bauer, Bauman, Baumont, Becker, Becla, Beldica,
  Bellavia, Bianco, Biswas, Blanc, Blazek, Blandford, Bloom, Bogart, Bond,
  Borgland, Borne, Bosch, Boutigny, Brackett, Bradshaw, Nielsen~Brandt, Brown,
  Bullock, Burchat, Burke, Cagnoli, Calabrese, Callahan, Callen,
  Chandrasekharan, Charles-Emerson, Chesley, Cheu, Chiang, Chiang, Chirino,
  Chow, Ciardi, Claver, Cohen-Tanugi, Cockrum, Coles, Connolly, Cook, Cooray,
  Covey, Cribbs, Cui, Cutri, Daly, Daniel, Daruich, Daubard, Daues, Dawson,
  Delgado, Dellapenna, de~Peyster, de~Val-Borro, Digel, Doherty, Dubois,
  Dubois-Felsmann, Durech, Economou, Eracleous, Ferguson, Figueroa,
  Fisher-Levine, Focke, Foss, Frank, Freemon, Gangler, Gawiser, Geary, Gee,
  Geha, Gessner, Gibson, Gilmore, Glanzman, Glick, Goldina, Goldstein,
  Goodenow, Graham, Gressler, Gris, Guy, Guyonnet, Haller, Harris, Hascall,
  Haupt, Hernandez, Herrmann, Hileman, Hoblitt, Hodgson, Hogan, Huang, Huffer,
  Ingraham, Innes, Jacoby, Jain, Jammes, Jee, Jenness, Jernigan,
  Jevremovi{\'c}, Johns, Johnson, Johnson, Jones, Juramy-Gilles, Juri{\'c},
  Kalirai, Kallivayalil, Kalmbach, Kantor, Karst, Kasliwal, Kelly, Kessler,
  Kinnison, Kirkby, Knox, Kotov, Krabbendam, Krughoff, Kub{\'a}nek, Kuczewski,
  Kulkarni, Ku, Kurita, Lage, Lambert, Lange, Langton, Le~Guillou, Levine,
  Liang, Lim, Lintott, Long, Lopez, Lotz, Lupton, Lust, MacArthur, Mahabal,
  Mandelbaum, Marsh, Marshall, Marshall, May, McKercher, McQueen, Meyers,
  Migliore, Miller, Mills, Miraval, Moeyens, Monet, Moniez, Monkewitz,
  Montgomery, Mueller, Muller, Mu{\~n}oz~Arancibia, Neill, Newbry, Nief,
  Nomerotski, Nordby, O'Connor, Oliver, Olivier, Olsen, O'Mullane, Ortiz,
  Osier, Owen, Pain, Palecek, Parejko, Parsons, Pease, Peterson, Peterson,
  Petravick, Libby~Petrick, Petry, Pierfederici, Pietrowicz, Pike, Pinto,
  Plante, Plate, Price, Prouza, Radeka, Rajagopal, \& Rasmussen}]{Ivezic08}
Ivezi{\'c}, {\v Z}., Kahn, S.~M., Tyson, J.~A., {et~al.} 2008, arXiv.org,
  arXiv:0805.2366

\bibitem[{Jiang {et~al.}(2018)Jiang, Wu, Bian, Chiang, Ho, Shen, Zheng, Bailey,
  Blanc, Crane, Fan, Mateo, Olszewski, Oyarz{\'u}n, Wang, \&
  Wu}]{2018NatAs...2..962J}
Jiang, L., Wu, J., Bian, F., {et~al.} 2018, Nature Astronomy, 2, 962

\bibitem[{Juri{\'c} {et~al.}(2015)Juri{\'c}, Kantor, Lim, Lupton,
  Dubois-Felsmann, Jenness, Axelrod, Allsman, AlSayyad, Alt, Armstrong, Basney,
  Becker, Becla, Bickerton, Biswas, Bosch, Boutigny, Kind, Ciardi, Connolly,
  Daniel, Daues, Economou, Chiang, Fausti, Fisher-Levine, Freemon, Gee, Gris,
  Hernandez, Hoblitt, Jammes, Jones, Kalmbach, Kasliwal, Krughoff, Lang, Lurie,
  Lust, Mullally, MacArthur, Melchior, Moeyens, Nidever, Owen, Parejko,
  Peterson, Petravick, Pietrowicz, Price, Reiss, Shaw, Sick, Slater, Strauss,
  Sullivan, Swinbank, Van~Dyk, Withers, Yoachim, \& Project}]{Juric15}
Juri{\'c}, M., Kantor, J., Lim, K.~T., {et~al.} 2015, arXiv.org

\bibitem[{Kawanomoto {et~al.}(2018)Kawanomoto, Uraguchi, Komiyama, Miyazaki,
  Furusawa, Finet, Hattori, Wang, Yasuda, \& Suzuki}]{Kawanomoto18}
Kawanomoto, S., Uraguchi, F., Komiyama, Y., {et~al.} 2018, \pasj, 70, 66

\bibitem[{Kennicutt(1998)}]{1998ARA&A..36..189K}
Kennicutt, R. C.~J. 1998, \araa, 36, 189

\bibitem[{Komiyama {et~al.}(2018)Komiyama, Obuchi, Nakaya, Kamata, Kawanomoto,
  Utsumi, Miyazaki, Uraguchi, Furusawa, Morokuma, Uchida, Miyatake, Mineo,
  Fujimori, Aihara, Karoji, Gunn, \& Wang}]{Komiyama18}
Komiyama, Y., Obuchi, Y., Nakaya, H., {et~al.} 2018, \pasj, 70, S2

\bibitem[{Konno {et~al.}(2016)Konno, Ouchi, Nakajima, Duval, Kusakabe, Ono, \&
  Shimasaku}]{2016ApJ...823...20K}
Konno, A., Ouchi, M., Nakajima, K., {et~al.} 2016, \apj, 823, 20

\bibitem[{Kovac {et~al.}(2010)Kovac, Lilly, Knobel, Bolzonella, Iovino,
  Carollo, Scarlata, Sargent, Cucciati, Zamorani, Pozzetti, Tasca, Scodeggio,
  Kampczyk, Peng, Oesch, Zucca, Finoguenov, Contini, Kneib, Le~Fevre, Mainieri,
  Renzini, Bardelli, Bongiorno, Caputi, Coppa, de~la Torre, de~Ravel,
  Franzetti, Garilli, Lamareille, Le~Borgne, Le~Brun, Maier, Mignoli,
  Pell{\'o}, P{\'e}rez-Montero, Ricciardelli, Silverman, Tanaka, Tresse,
  Vergani, Abbas, Bottini, Cappi, Cassata, Cimatti, Fumana, Guzzo, Koekemoer,
  Leauthaud, Maccagni, Marinoni, McCracken, Memeo, Meneux, Porciani,
  Scaramella, \& Scoville}]{2010ApJ...718...86K}
Kovac, K., Lilly, S.~J., Knobel, C., {et~al.} 2010, \apj, 718, 86

\bibitem[{Koyama {et~al.}(2013)Koyama, Kodama, Tadaki, Hayashi, Tanaka, Smail,
  Tanaka, \& Kurk}]{2013MNRAS.428.1551K}
Koyama, Y., Kodama, T., Tadaki, K.-i., {et~al.} 2013, \mnras, 428, 1551

\bibitem[{Krishnan {et~al.}(2017)Krishnan, Hatch, Almaini, Kocevski, Cooke,
  Hartley, Hasinger, Maltby, Muldrew, \& Simpson}]{Krishnan17}
Krishnan, C., Hatch, N.~A., Almaini, O., {et~al.} 2017, \mnras, 470, 2170

\bibitem[{Kubo {et~al.}(2019)Kubo, Toshikawa, Kashikawa, Chiang, Overzier,
  Uchiyama, Clements, Alexander, Matsuda, Kodama, Ono, Goto, Cheng, \&
  Ito}]{2019ApJ...887..214K}
Kubo, M., Toshikawa, J., Kashikawa, N., {et~al.} 2019, \apj, 887, 214

\bibitem[{Kurk {et~al.}(2009)Kurk, Cimatti, Zamorani, Halliday, Mignoli,
  Pozzetti, Daddi, Rosati, Dickinson, Bolzonella, Cassata, Renzini,
  Franceschini, Rodighiero, \& Berta}]{2009A&A...504..331K}
Kurk, J., Cimatti, A., Zamorani, G., {et~al.} 2009, \aap, 504, 331

\bibitem[{Lee {et~al.}(2014)Lee, Hennawi, White, Croft, \&
  Ozbek}]{2014ApJ...788...49L}
Lee, K.-G., Hennawi, J.~F., White, M., Croft, R. A.~C., \& Ozbek, M. 2014,
  \apj, 788, 49

\bibitem[{Lee {et~al.}(2016)Lee, Hennawi, White, Prochaska, Font-Ribera,
  Schlegel, Rich, Suzuki, Stark, Le~Fevre, Nugent, Salvato, \&
  Zamorani}]{2016ApJ...817..160L}
Lee, K.-G., Hennawi, J.~F., White, M., {et~al.} 2016, \apj, 817, 160

\bibitem[{Lehmer {et~al.}(2009)Lehmer, Alexander, Geach, Smail, Basu-Zych,
  Bauer, Chapman, Matsuda, Scharf, Volonteri, \& Yamada}]{2009ApJ...691..687L}
Lehmer, B.~D., Alexander, D.~M., Geach, J.~E., {et~al.} 2009, \apj, 691, 687

\bibitem[{Lemaux {et~al.}(2018)Lemaux, Le~Fevre, Cucciati, Ribeiro, Tasca,
  Zamorani, Ilbert, Thomas, Bardelli, Cassata, Hathi, Pforr, Smol{\v c}i{\'c},
  Delvecchio, Novak, Berta, McCracken, Koekemoer, Amorin, Garilli, Maccagni,
  Schaerer, \& Zucca}]{2018A&A...615A..77L}
Lemaux, B.~C., Le~Fevre, O., Cucciati, O., {et~al.} 2018, \aap, 615, A77

\bibitem[{Lidman {et~al.}(2012)Lidman, Suherli, Muzzin, Wilson, Demarco,
  Brough, Rettura, Cox, DeGroot, Yee, Gilbank, Hoekstra, Balogh, Ellingson,
  Hicks, Nantais, Noble, Lacy, Surace, \& Webb}]{2012MNRAS.427..550L}
Lidman, C., Suherli, J., Muzzin, A., {et~al.} 2012, \mnras, 427, 550

\bibitem[{Lin {et~al.}(2017)Lin, Hsieh, Lin, Oguri, Chen, Tanaka, Chiu, Huang,
  Kodama, Leauthaud, More, Nishizawa, Bundy, Lin, \&
  Miyazaki}]{2017ApJ...851..139L}
Lin, Y.-T., Hsieh, B.-C., Lin, S.-C., {et~al.} 2017, \apj, 851, 139

\bibitem[{Long {et~al.}(2020)Long, Cooray, Ma, Casey, Wardlow, Nayyeri, Ivison,
  Farrah, \& Dannerbauer}]{Long:2020vn}
Long, A.~S., Cooray, A., Ma, J., {et~al.} 2020, arXiv.org

\bibitem[{Lovell {et~al.}(2018)Lovell, Thomas, \&
  Wilkins}]{2018MNRAS.474.4612L}
Lovell, C.~C., Thomas, P.~A., \& Wilkins, S.~M. 2018, \mnras, 474, 4612

\bibitem[{Lovell {et~al.}(2020)Lovell, Vijayan, Thomas, Wilkins, Barnes,
  Irodotou, \& Roper}]{Lovell:2020wz}
Lovell, C.~C., Vijayan, A.~P., Thomas, P.~A., {et~al.} 2020, arXiv.org

\bibitem[{Macuga {et~al.}(2019)Macuga, Martini, Miller, Brodwin, Hayashi,
  Kodama, Koyama, Overzier, Shimakawa, Tadaki, \& Tanaka}]{2019ApJ...874...54M}
Macuga, M., Martini, P., Miller, E.~D., {et~al.} 2019, \apj, 874, 54

\bibitem[{Madau \& Dickinson(2014)}]{2014ARA&A..52..415M}
Madau, P., \& Dickinson, M. 2014, \araa, 52, 415

\bibitem[{Marrone {et~al.}(2018)Marrone, Spilker, Hayward, Vieira, Aravena,
  Ashby, Bayliss, B{\'e}thermin, Brodwin, Bothwell, Carlstrom, Chapman, Chen,
  Crawford, Cunningham, De~Breuck, Fassnacht, Gonzalez, Greve, Hezaveh,
  Lacaille, Litke, Lower, Ma, Malkan, Miller, Morningstar, Murphy, Narayanan,
  Phadke, Rotermund, Sreevani, Stalder, Stark, Strandet, Tang, \&
  Wei{\ss}}]{2018Natur.553...51M}
Marrone, D.~P., Spilker, J.~S., Hayward, C.~C., {et~al.} 2018, \nat, 553, 51

\bibitem[{Meiksin(2006)}]{2006MNRAS.365..807M}
Meiksin, A. 2006, \mnras, 365, 807

\bibitem[{Meurer {et~al.}(1999)Meurer, Heckman, \& Calzetti}]{Meurer99}
Meurer, G.~R., Heckman, T.~M., \& Calzetti, D. 1999, \apj, 521, 64

\bibitem[{Miller {et~al.}(2018)Miller, Chapman, Aravena, Ashby, Hayward,
  Vieira, Wei{\ss}, Babul, B{\'e}thermin, Bradford, Brodwin, Carlstrom, Chen,
  Cunningham, De~Breuck, Gonzalez, Greve, Harnett, Hezaveh, Lacaille, Litke,
  Ma, Malkan, Marrone, Morningstar, Murphy, Narayanan, Pass, Perry, Phadke,
  Rennehan, Rotermund, Simpson, Spilker, Sreevani, Stark, Strandet, \&
  Strom}]{2018Natur.556..469M}
Miller, T.~B., Chapman, S.~C., Aravena, M., {et~al.} 2018, Nature, 556, 469

\bibitem[{Miyazaki {et~al.}(2018)Miyazaki, Komiyama, Kawanomoto, Doi, Furusawa,
  Hamana, Hayashi, Ikeda, Kamata, Karoji, Koike, Kurakami, Miyama, Morokuma,
  Nakata, Namikawa, Nakaya, Nariai, Obuchi, Oishi, Okada, Okura, Tait, Takata,
  Tanaka, Tanaka, Terai, Tomono, Uraguchi, Usuda, Utsumi, Yamada, Yamanoi,
  Aihara, Fujimori, Mineo, Miyatake, Oguri, Uchida, Tanaka, Yasuda, Takada,
  Murayama, Nishizawa, Sugiyama, Chiba, Futamase, Wang, Chen, Ho, Liaw, Chiu,
  Ho, Lai, Lee, Jeng, Iwamura, Armstrong, Bickerton, Bosch, Gunn, Lupton,
  Loomis, Price, Smith, Strauss, Turner, Suzuki, Miyazaki, Muramatsu, Yamamoto,
  Endo, Ezaki, Ito, Kawaguchi, Sofuku, Taniike, Akutsu, Dojo, Kasumi, Matsuda,
  Imoto, Miwa, Suzuki, Takeshi, \& Yokota}]{2018PASJ...70S...1M}
Miyazaki, S., Komiyama, Y., Kawanomoto, S., {et~al.} 2018, \pasj, 70, S1

\bibitem[{Muldrew {et~al.}(2015)Muldrew, Hatch, \& Cooke}]{2015MNRAS.452.2528M}
Muldrew, S.~I., Hatch, N.~A., \& Cooke, E.~A. 2015, \mnras, 452, 2528

\bibitem[{Muzzin {et~al.}(2009)Muzzin, Wilson, Yee, Hoekstra, Gilbank, Surace,
  Lacy, Blindert, Majumdar, Demarco, Gardner, Gladders, \&
  Lonsdale}]{2009ApJ...698.1934M}
Muzzin, A., Wilson, G., Yee, H. K.~C., {et~al.} 2009, \apj, 698, 1934

\bibitem[{Nantais {et~al.}(2016)Nantais, van~der Burg, Lidman, Demarco, Noble,
  Wilson, Muzzin, Foltz, DeGroot, \& Cooper}]{2016A&A...592A.161N}
Nantais, J.~B., van~der Burg, R. F.~J., Lidman, C., {et~al.} 2016, \aap, 592,
  A161

\bibitem[{Newman {et~al.}(2014)Newman, Ellis, Andreon, Treu, Raichoor, \&
  Trinchieri}]{2014ApJ...788...51N}
Newman, A.~B., Ellis, R.~S., Andreon, S., {et~al.} 2014, \apj, 788, 51

\bibitem[{Oguri {et~al.}(2018)Oguri, Lin, Lin, Nishizawa, More, More, Hsieh,
  Medezinski, Miyatake, Jian, Lin, Takada, Okabe, Speagle, Coupon, Leauthaud,
  Lupton, Miyazaki, Price, Tanaka, Chiu, Komiyama, Okura, Tanaka, \&
  Usuda}]{2018PASJ...70S..20O}
Oguri, M., Lin, Y.-T., Lin, S.-C., {et~al.} 2018, \pasj, 70, S20

\bibitem[{Ono {et~al.}(2018)Ono, Ouchi, Harikane, Toshikawa, Rauch, Yuma,
  Sawicki, Shibuya, Shimasaku, Oguri, Willott, Akhlaghi, Akiyama, Coupon,
  Kashikawa, Komiyama, Konno, Lin, Matsuoka, Miyazaki, Nagao, Nakajima,
  Silverman, Tanaka, Taniguchi, \& Wang}]{Ono2018}
Ono, Y., Ouchi, M., Harikane, Y., {et~al.} 2018, \pasj, 70, S10

\bibitem[{Onoue {et~al.}(2018)Onoue, Kashikawa, Uchiyama, Akiyama, Harikane,
  Imanishi, Komiyama, Matsuoka, Nagao, Nishizawa, Oguri, Ouchi, Tanaka, Toba,
  \& Toshikawa}]{2018PASJ...70S..31O}
Onoue, M., Kashikawa, N., Uchiyama, H., {et~al.} 2018, \pasj, 70, S31

\bibitem[{Ouchi {et~al.}(2005)Ouchi, Shimasaku, Akiyama, Sekiguchi, Furusawa,
  Okamura, Kashikawa, Iye, Kodama, Saito, Sasaki, Simpson, Takata, Yamada,
  Yamanoi, Yoshida, \& Yoshida}]{2005ApJ...620L...1O}
Ouchi, M., Shimasaku, K., Akiyama, M., {et~al.} 2005, \apj, 620, L1

\bibitem[{Overzier(2016)}]{Overzier16}
Overzier, R.~A. 2016, \aapr, 24, 14

\bibitem[{Overzier {et~al.}(2008)Overzier, Bouwens, Cross, Venemans, Miley,
  Zirm, Benitez, Blakeslee, Coe, Demarco, Ford, Homeier, Illingworth, Kurk,
  Martel, Mei, Oliveira, R{\"o}ttgering, Tsvetanov, \&
  Zheng}]{2008ApJ...673..143O}
Overzier, R.~A., Bouwens, R.~J., Cross, N. J.~G., {et~al.} 2008, \apj, 673, 143

\bibitem[{Rowe {et~al.}(2015)Rowe, Jarvis, Mandelbaum, Bernstein, Bosch, Simet,
  Meyers, Kacprzak, Nakajima, Zuntz, Miyatake, Dietrich, Armstrong, Melchior,
  \& Gill}]{2015A&C....10..121R}
Rowe, B. T.~P., Jarvis, M., Mandelbaum, R., {et~al.} 2015, Astronomy and
  Computing, 10, 121

\bibitem[{Salpeter(1955)}]{1955ApJ...121..161S}
Salpeter, E.~E. 1955, \apj, 121, 161

\bibitem[{Schechter \& Press(1976)}]{1976ApJ...203..557S}
Schechter, P., \& Press, W.~H. 1976, \apj, 203, 557

\bibitem[{Schlegel {et~al.}(1998)Schlegel, Finkbeiner, \& Davis}]{Schlegel98}
Schlegel, D.~J., Finkbeiner, D.~P., \& Davis, M. 1998, \apj, 500, 525

\bibitem[{S{\'e}rsic(1963)}]{Sersic63}
S{\'e}rsic, J.~L. 1963, Boletin de la Asociacion Argentina de Astronomia, 6, 41

\bibitem[{Shi {et~al.}(2019{\natexlab{a}})Shi, Lee, Dey, Huang, Malavasi, Hung,
  Inami, Ashby, Duncan, Xue, Reddy, Hong, Jannuzi, Cooper, Gonzalez,
  R{\"o}ttgering, Best, \& Tasse}]{2019ApJ...871...83S}
Shi, K., Lee, K.-S., Dey, A., {et~al.} 2019{\natexlab{a}}, \apj, 871, 83

\bibitem[{Shi {et~al.}(2019{\natexlab{b}})Shi, Huang, Lee, Toshikawa, Bowen,
  Malavasi, Lemaux, Cucciati, Le~Fevre, \& Dey}]{2019ApJ...879....9S}
Shi, K., Huang, Y., Lee, K.-S., {et~al.} 2019{\natexlab{b}}, \apj, 879, 9

\bibitem[{Shibuya {et~al.}(2015)Shibuya, Ouchi, \&
  Harikane}]{2015ApJS..219...15S}
Shibuya, T., Ouchi, M., \& Harikane, Y. 2015, \apjs, 219, 15

\bibitem[{Shimakawa {et~al.}(2018)Shimakawa, Kodama, Hayashi, Prochaska,
  Tanaka, Cai, Suzuki, Tadaki, \& Koyama}]{2018MNRAS.473.1977S}
Shimakawa, R., Kodama, T., Hayashi, M., {et~al.} 2018, \mnras, 473, 1977

\bibitem[{Sohn {et~al.}(2018)Sohn, Geller, Rines, Hwang, Utsumi, \&
  Diaferio}]{2018ApJ...856..172S}
Sohn, J., Geller, M.~J., Rines, K.~J., {et~al.} 2018, \apj, 856, 172

\bibitem[{Song {et~al.}(2016)Song, Finkelstein, Ashby, Grazian, Lu, Papovich,
  Salmon, Somerville, Dickinson, Duncan, Faber, Fazio, Ferguson, Fontana, Guo,
  Hathi, Lee, Merlin, \& Willner}]{Song16}
Song, M., Finkelstein, S.~L., Ashby, M. L.~N., {et~al.} 2016, \apj, 825, 5

\bibitem[{Speagle {et~al.}(2014)Speagle, Steinhardt, Capak, \&
  Silverman}]{Speagle14}
Speagle, J.~S., Steinhardt, C.~L., Capak, P.~L., \& Silverman, J.~D. 2014,
  \apjs, 214, 15

\bibitem[{Stark {et~al.}(2015)Stark, Font-Ribera, White, \&
  Lee}]{2015MNRAS.453.4311S}
Stark, C.~W., Font-Ribera, A., White, M., \& Lee, K.-G. 2015, \mnras, 453, 4311

\bibitem[{Steidel {et~al.}(1998)Steidel, Adelberger, Dickinson, Giavalisco,
  Pettini, \& Kellogg}]{Steidel98}
Steidel, C.~C., Adelberger, K.~L., Dickinson, M., {et~al.} 1998, \apj, 492, 428

\bibitem[{Steidel {et~al.}(2005)Steidel, Adelberger, Shapley, Erb, Reddy, \&
  Pettini}]{2005ApJ...626...44S}
Steidel, C.~C., Adelberger, K.~L., Shapley, A.~E., {et~al.} 2005, \apj, 626, 44

\bibitem[{Suchyta {et~al.}(2016)Suchyta, Huff, Aleksi{\'c}, Melchior, Jouvel,
  MacCrann, Ross, Crocce, Gaztanaga, Honscheid, Leistedt, Peiris, Rykoff,
  Sheldon, Abbott, Abdalla, Allam, Banerji, Benoit-L{\'e}vy, Bertin, Brooks,
  Burke, Carnero~Rosell, Carrasco~Kind, Carretero, Cunha, D'Andrea, da~Costa,
  DePoy, Desai, Diehl, Dietrich, Doel, Eifler, Estrada, Evrard, Flaugher,
  Fosalba, Frieman, Gerdes, Gruen, Gruendl, James, Jarvis, Kuehn, Kuropatkin,
  Lahav, Lima, Maia, March, Marshall, Miller, Miquel, Neilsen, Nichol, Nord,
  Ogando, Percival, Reil, Roodman, Sako, Sanchez, Scarpine, Sevilla-Noarbe,
  Smith, Soares-Santos, Sobreira, Swanson, Tarle, Thaler, Thomas, Vikram,
  Walker, Wechsler, Zhang, \& des Collaboration}]{2016MNRAS.457..786S}
Suchyta, E., Huff, E.~M., Aleksi{\'c}, J., {et~al.} 2016, \mnras, 457, 786

\bibitem[{Tanaka {et~al.}(2019)Tanaka, Valentino, Toft, Onodera, Shimakawa,
  Ceverino, Faisst, Gallazzi, G{\'o}mez-Guijarro, Kubo, Magdis, Steinhardt,
  Stockmann, Yabe, \& Zabl}]{2019ApJ...885L..34T}
Tanaka, M., Valentino, F., Toft, S., {et~al.} 2019, \apjl, 885, L34

\bibitem[{Thomas {et~al.}(2005)Thomas, Maraston, Bender, \& Mendes~de
  Oliveira}]{Thomas05}
Thomas, D., Maraston, C., Bender, R., \& Mendes~de Oliveira, C. 2005, \apj,
  621, 673

\bibitem[{Tomczak {et~al.}(2016)Tomczak, Quadri, Tran, Labb{\'e}, Straatman,
  Papovich, Glazebrook, Allen, Brammer, Cowley, Dickinson, Elbaz, Inami,
  Kacprzak, Morrison, Nanayakkara, Persson, Rees, Salmon, Schreiber, Spitler,
  \& Whitaker}]{2016ApJ...817..118T}
Tomczak, A.~R., Quadri, R.~F., Tran, K.-V.~H., {et~al.} 2016, \apj, 817, 118

\bibitem[{Tomczak {et~al.}(2017)Tomczak, Lemaux, Lubin, Gal, Wu, Holden,
  Kocevski, Mei, Pelliccia, Rumbaugh, \& Shen}]{2017MNRAS.472.3512T}
Tomczak, A.~R., Lemaux, B.~C., Lubin, L.~M., {et~al.} 2017, \mnras, 472, 3512

\bibitem[{Toshikawa {et~al.}(2020)Toshikawa, Malkan, Kashikawa, Overzier,
  Uchiyama, Ota, Ishikawa, \& Ito}]{2020ApJ...888...89T}
Toshikawa, J., Malkan, M.~A., Kashikawa, N., {et~al.} 2020, The Astrophysical
  Journal, 888, 89

\bibitem[{Toshikawa {et~al.}(2012)Toshikawa, Kashikawa, Ota, Morokuma, Shibuya,
  Hayashi, Nagao, Jiang, Malkan, Egami, Shimasaku, Motohara, \&
  Ishizaki}]{Toshikawa12}
Toshikawa, J., Kashikawa, N., Ota, K., {et~al.} 2012, \apj, 750, 137

\bibitem[{Toshikawa {et~al.}(2016)Toshikawa, Kashikawa, Overzier, Malkan,
  Furusawa, Ishikawa, Onoue, Ota, Tanaka, Niino, \& Uchiyama}]{Toshikawa16}
Toshikawa, J., Kashikawa, N., Overzier, R.~A., {et~al.} 2016, \apj, 826, 114

\bibitem[{Toshikawa {et~al.}(2018)Toshikawa, Uchiyama, Kashikawa, Ouchi,
  Overzier, Ono, Harikane, Ishikawa, Kodama, Matsuda, Lin, Onoue, Tanaka,
  Nagao, Akiyama, Komiyama, Goto, \& Lee}]{Toshikawa18}
Toshikawa, J., Uchiyama, H., Kashikawa, N., {et~al.} 2018, \pasj, 70, S12

\bibitem[{Uchiyama {et~al.}(2018)Uchiyama, Toshikawa, Kashikawa, Overzier,
  Chiang, Marinello, Tanaka, Niino, Ishikawa, Onoue, Ichikawa, Akiyama, Coupon,
  Harikane, Imanishi, Kodama, Komiyama, Lee, Lin, Miyazaki, Nagao, Nishizawa,
  Ono, Ouchi, \& Wang}]{2018PASJ...70S..32U}
Uchiyama, H., Toshikawa, J., Kashikawa, N., {et~al.} 2018, \pasj, 70, S32

\bibitem[{Valentino {et~al.}(2020)Valentino, Tanaka, Davidzon, Toft,
  G{\'o}mez-Guijarro, Stockmann, Onodera, Brammer, Ceverino, Faisst, Gallazzi,
  Hayward, Ilbert, Kubo, Magdis, Selsing, Shimakawa, Sparre, Steinhardt, Yabe,
  \& Zabl}]{2020ApJ...889...93V}
Valentino, F., Tanaka, M., Davidzon, I., {et~al.} 2020, \apj, 889, 93

\bibitem[{van~der Burg {et~al.}(2010)van~der Burg, Hildebrandt, \&
  Erben}]{VanderBurg2010}
van~der Burg, R. F.~J., Hildebrandt, H., \& Erben, T. 2010, \aap, 523, A74

\bibitem[{van~der Burg {et~al.}(2018)van~der Burg, McGee, Aussel, Dahle,
  Arnaud, Pratt, \& Muzzin}]{2018A&A...618A.140V}
van~der Burg, R. F.~J., McGee, S., Aussel, H., {et~al.} 2018, \aap, 618, A140

\bibitem[{van~der Burg {et~al.}(2013)van~der Burg, Muzzin, Hoekstra, Lidman,
  Rettura, Wilson, Yee, Hildebrandt, Marchesini, Stefanon, Demarco, \&
  Kuijken}]{vanderBurg:2013gl}
van~der Burg, R. F.~J., Muzzin, A., Hoekstra, H., {et~al.} 2013, \aap, 557, A15

\bibitem[{Venemans {et~al.}(2002)Venemans, Kurk, Miley, R{\"o}ttgering, van
  Breugel, Carilli, De~Breuck, Ford, Heckman, McCarthy, \&
  Pentericci}]{2002ApJ...569L..11V}
Venemans, B.~P., Kurk, J.~D., Miley, G.~K., {et~al.} 2002, \apj, 569, L11

\bibitem[{Venemans {et~al.}(2007)Venemans, R{\"o}ttgering, Miley, van Breugel,
  De~Breuck, Kurk, Pentericci, Stanford, Overzier, Croft, \& Ford}]{Venemans07}
Venemans, B.~P., R{\"o}ttgering, H. J.~A., Miley, G.~K., {et~al.} 2007, \aap,
  461, 823

\bibitem[{Vulcani {et~al.}(2013)Vulcani, Poggianti, Oemler, Dressler,
  Arag{\'o}n-Salamanca, De~Lucia, Moretti, Gladders, Abramson, \&
  Halliday}]{2013A&A...550A..58V}
Vulcani, B., Poggianti, B.~M., Oemler, A., {et~al.} 2013, \aap, 550, A58

\bibitem[{Wang {et~al.}(2016)Wang, Elbaz, Daddi, Finoguenov, Liu, Schreiber,
  Mart{\'\i}n, Strazzullo, Valentino, van~der Burg, Zanella, Ciesla, Gobat,
  Le~Brun, Pannella, Sargent, Shu, Tan, Cappelluti, \&
  Li}]{2016ApJ...828...56W}
Wang, T., Elbaz, D., Daddi, E., {et~al.} 2016, \apj, 828, 56

\bibitem[{Wang {et~al.}(2019)Wang, Schreiber, Elbaz, Yoshimura, Kohno, Shu,
  Yamaguchi, Pannella, Franco, Huang, Lim, \& Wang}]{2019Natur.572..211W}
Wang, T., Schreiber, C., Elbaz, D., {et~al.} 2019, \nat, 572, 211

\bibitem[{Wilson {et~al.}(2009)Wilson, Muzzin, Yee, Lacy, Surace, Gilbank,
  Blindert, Hoekstra, Majumdar, Demarco, Gardner, Gladders, \&
  Lonsdale}]{2009ApJ...698.1943W}
Wilson, G., Muzzin, A., Yee, H. K.~C., {et~al.} 2009, \apj, 698, 1943

\bibitem[{Yoshida {et~al.}(2006)Yoshida, Shimasaku, Kashikawa, Ouchi, Okamura,
  Ajiki, Akiyama, Ando, Aoki, Doi, Furusawa, Hayashino, Iwamuro, Iye, Karoji,
  Kobayashi, Kodaira, Kodama, Komiyama, Malkan, Matsuda, Miyazaki, Mizumoto,
  Morokuma, Motohara, Murayama, Nagao, Nariai, Ohta, Sasaki, Sato, Sekiguchi,
  Shioya, Tamura, Taniguchi, Umemura, Yamada, \& Yasuda}]{2006ApJ...653..988Y}
Yoshida, M., Shimasaku, K., Kashikawa, N., {et~al.} 2006, \apj, 653, 988

\end{thebibliography}

\end{document}